# Advances of Machine Learning in Materials Science: Ideas and Techniques

Sue Sin Chong[1#], Yi Sheng Ng[1#], Hui-Qiong Wang[1,2*], and Jin-Cheng Zheng[1,2*]

[1]Department of New Energy Science and Engineering, Xiamen University Malaysia, Sepang 43900, Malaysia

[2]Engineering Research Center of Micro-nano Optoelectronic Materials and Devices, Ministry of Education; Fujian Key Laboratory of Semiconductor Materials and Applications, CI Center for OSED, and Department of Physics, Xiamen University, Xiamen 361005, China

[*]Corresponding Author: hqwang@xmu.edu.cn, jczheng@xmu.edu.cn
[#] These two authors contribute equally.

## Abstract

In this big data era, the use of large dataset in conjunction with machine learning (ML) has been increasingly popular in both industry and academia. In recent times, the field of materials science is also undergoing a big data revolution, with large database and repositories appearing everywhere. Traditionally, materials science is a trial-and-error field, in both the computational and experimental departments. With the advent of machine learning-based techniques, there has been a paradigm shift: materials can now be screened quickly using ML models and even generated based on materials with similar properties; ML has also quietly infiltrated many sub-disciplinary under materials science. However, ML remains relatively new to the field and is expanding its wing quickly. There are a plethora of readily-available big data architectures and abundance of ML models and software; The call to integrate all these elements in a comprehensive research procedure is becoming an important direction of material science research. In this review, we attempt to provide an introduction and reference of ML to materials scientists, covering as much as possible the commonly used methods and applications, and discussing the future possibilities.

## Contents

## 1 Introduction

Commonly recognized as the fourth paradigm of science [1-4], machine learning (ML) has played a crucial role in the development of the data-driven scientific process, shaping the changes in experimental methodology, from measurement to data analysis, assisting the proving of mathematical theorem, and finding new discoveries in areas that were once deemed impossible. In the field of computational material science, the methods have been enriched by the efficient high-throughput ML-aided simulation/data generation and data-driven discovery [5, 6]. In experiments and materials synthesis, the advances in ML have helped researchers to efficiently analyze the data and identify hidden features within the large dataset [7-9].

Discovery of new internal logics, patterns, or rules [10-13], and the study of complex systems, including nanostructures [14-29], alloys [30-35], superlattices [22,36-38], surfaces [39-41], and interfaces [40,42-46] , as well as from materials to devices[47-48], are typical research topics in materials science. These areas could be addressed according to user specifications [49-50] by leveraging ML and big data statistical methods [51], which have advanced to a stage where users can utilize them to achieve large and complicated objectives with complex models. By breaking down broad objectives into smaller tasks, corresponding ML algorithms and objectives that are suitable can be identified and applied.

There are numerous comprehensive surveys on ML in material science [52-60]. In this review, we focus on the application of ML in material science, discussing the recent advances in ML, illustrating the basic principles of applying ML in materials science, and summarizing the current applications and briefly introducing the ML algorithms involved.

## 2 Basics on Machine Learning

ML has a long history [61-67]. However, it has only returned to the spotlight recently due to the compounding ability it has gained from the surge in big data and improving data infrastructure and computing power. Stemmed from statistical learning, ML has gained huge successes and popularity in many other tasks and has far-reaching influences in many fields, including physics, chemistry and material science. In this section, the basic ideas and concepts in ML and essential milestones in its illustrious history are covered.

ML can be broadly defined as computational methods using experience (available past information) to improve future performance or to make accurate predictions [68]. Typical ML methods involve three parts: the inputs (previously obtained data), outputs (predictions), and algorithms. The sample size (sample complexity) and the time & space complexity of algorithms are crucial for ML [68]. Therefore, the ML techniques are different from conventional methods such as experimental measurements or computer simulations, but are related to data analysis and statistics. In more general terms, ML techniques are data-driven methods, which combines the fundamental concepts of computer science with ideas from statistics, probability, and optimization. ML can be integrated with other disciplines, resulting in multi-discipline techiques such as quantum ML or physics-informed neural networks, materials science-based learning.

In ML, the "standard" or conventional learning tasks have been extensively studied, which include classification, regression, ranking, clustering, and dimensionality reduction or manifold learning [68]. The problems related to the above tasks are listed in Figure 1. The definitions and terminology commonly used in ML for different learning stages are listed in Figure 2. The typical stages of a learning process are also shown in Figure 3, which can be briefly described as follows: with a given collection of labeled examples, one can firstly divide the data/samples into three groups, namely, training samples, validation data and test samples, then the relevant features associated to the desired properties are chosen, which are next used to train the pre-determined learning algorithm. This is done by adjusting the hyperparameters $\Theta$ in order to ensure that the hypothesis $\Theta_0$ has the best performance on the validation sample. Typical learning scenarios include supervised learning, unsupervised learning, semi-supervised learning, transductive inference, on-line learning, reinforcement learning, active learning, and more other complex learning scenarios. Different from traditional data analysis, ML is fundamentally about generalization [68]. Spectacularly, the neural network-based ML is able to approximate functions in a very high dimension with unprecedented efficiency and accuracy [2], and therefore it can be used for complex tasks in a wide-range of applications.

| Classification | • Problem: Assigning a category to each item. |
| Regression | • Problem: Predicting a real value for each item. |
| Ranking | • Problem: Learning to order items according to some criterion. |
| Clustering | • Problem: Partitioning a set of items into homogeneous subsets. |
| Dimensionality reduction or manifold learning | • Problem: Transforming an initial representation of items into a lower dimensional representation while preserving some properties of the initial representation. |

Figure 1. List of the conventional machine learning tasks and the problems tackled [68].

| Examples | • Items or instances of data used for learning or evaluation. |
|---|---|
| Features | • The set of attributes, often represented as a vector, associated to an example. |
| Labels | • Values or categories assigned to examples. |
| Hyperparameters | • Free parameters that are not determined by the learning algorithm, but rather specified as inputs to the learning algorithm. |
| Training sample | • Examples used to train a learning algorithm. |
| Validation sample | • Examples used to tune the parameters of a learning algorithm when working with labeled data. |
| Test sample | • Examples used to evaluate the performance of a learning algorithm. |
| Loss function | • A function that measures the difference, or loss, between a predicted label and a true label. |
| Hypothesis set | • A set of functions mapping features (feature vectors) to the set of labels. |

Figure 2. List of typical machine learning terminologies [68].

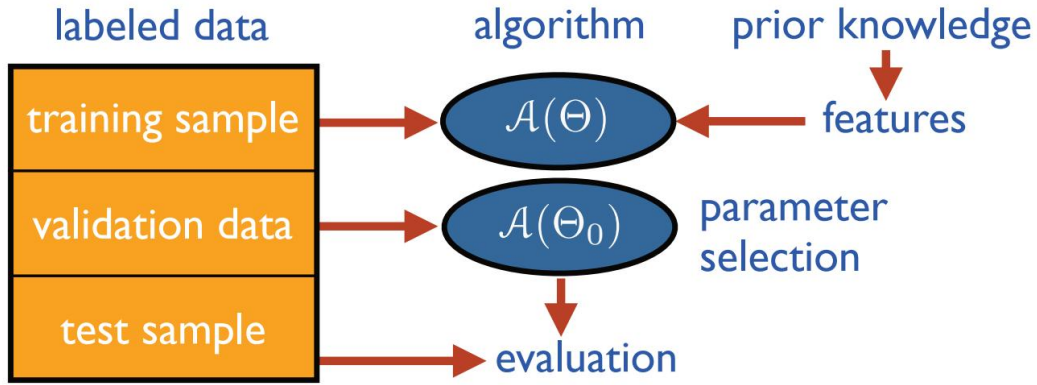

Figure 3. Illustration of the typical stages of a learning process. [68]

## 3  Recent Progress in Machine Learning

Recently, the ML community has seen breakthroughs in many traditional AI tasks and classical challenging scientific tasks. This leap of improvement is being powered by both the new grounds in the underlying theory, the overall implementation and architecture, and the massive surge in data and data infrastructures. This section covers the advancement of ideas from various areas of applications of Artificial Intelligence -Natural Language Processing (NLP), Computer Vision (CV), Reinforcement Learning (RL), Explainability Artificial Intellignce (XAI), etc.

### 3.1  Classical Machine Learning Application Areas

In the field of natural language processing and understanding, ML models have made huge progress with Attention Transformer networks [69] and pre-training techniques. SuperGLUE [70] is a natural language understanding benchmark consisting of many tasks, which requires in-depth understanding of short proses and sentences. With superhuman performances at SuperGLUE benchmarks, it has been demonstrated that ML is able to model both understanding of natural language and generation of relevant natural language in context. The technique that has led to this leap in performance is pre-training [71], which refers to "training a model with one task to help it form parameters that can be used in other tasks". Prompt learning is a form of ML that works with large models, to learn knowledge from a language model simply by prompting the learnt model with various types of prompts. BERT-like models have also been extended to process data from realms outside natural language, like Programming Languages e.g. CodeBERT [72], and Images [73], and had been very successful in these realms too. Table 1 lists works relevant to several main ideas in machine learning for Natural Language Processing (NLP).

Table 1: *Natural Language Processing (NLP) Ideas, Techniques and Models*

| Ideas & Technique | Relevant Development and Models |
|---|---|
| Pre-Training | Ref.[71], Ref.[74], BLIP [75], Pretrained Transformers [76], Ref. [77] |
| Fine-Tuning | Ref.[78] Ref.[79] |
| Bidirectional Encoder | BERT [80], Albert [81], Robustly Optimized BERT pre-training Approach (RoBERTa) [82], CodeBERT [72], BeiT [73] |
| Transformer | Ref.[69] Ref.[83] Ref.[84], Transfomer Memory as Search Index [85] |
| Attention Prompt | Ref.[86] AutoPrompt [87], OpenPrompt [88] |
| Learning Extra Huge Models | Open Pretrained Transformer (OPT 175B) [89], Jurassic-1 [90], Generative Pre-trained Transformer 3 (GPT-3) [91], CLD-3 [92] |
| End-to-end Model | Word2Vec [93], Global Vectors for Word Representation (GLoVE) [94], Context2Vec [95], Structure2Vec [96], Driver2Vec [97], wav2Vec [98] |

Unsupervised learning has made strides in computer vision tasks, with models being able to identify subject in video, or identify poses of objects from point cloud in a video without labels. In the area of unsupervised learning for time-series data, ML models are able to effectively identify features from time-series data for both classification and prediction. Recently, there has been many work that extends the use of transformers to the characterization and prediction of extra-long timeseries sequences, Informer [99], Longformer [100]. Table 2 lists works relevant to several main ideas in ML for Computer Vision (CV).

Table 2: *Computer Vision (CV) Ideas, Techniques and References*

| Ideas & Techniques | Relevant Literature |
|---|---|
| Visual Models | Visual Transformer [101], Flamingo (LM) [102] |
| Image-Text Processing | CoCa [103], FuseDream [104], CLIP [105] |
| Convolutional Neural Network | DEtection Transformer (DETR)[106], LiT [107], Ref. [108] |
| Image Rendering | Dall-E [109], Review [110], Neural Radiance Field (NeRF) [111] |
| Point Cloud Reconstruction | PointNorm [112], Ref. [113], Residual MLP [114], Learning on Point Cloud [115] |

Reinforcement learning is the training of agents to make a sequence of reward-optimal decisions in an environment, often modelled as maximizing reward expectation in a partial Markov Decision Process (MDP). In reinforcement learning thrust, there is a huge improvement in the ability of state-of-the-art models to effectively navigate extra large search space to search for sequential actions to maximize task goals. Most notably, models like AlphaGo [116] and AlphaHoldem [117] have been able to navigate extra large search spaces with Monte Carlo search tree methods on latent representation of state space and action. Classical methods like State–Action–Reward–State–Action (SARSA) [118], Q-Learning [119], TD-Learning explores action space with a reward function, and learns a matching of state space to action, a policy or the q-value of the actions. In 2012, apprenticeship learning [120] initiated by Abbeel, proposes to define the architecture such that agent is able to learn directly from observations of a task being completed, instead of specifying the steps of a task. There is also a trend to integrate reinforcement learning with meta-learning in order to train multi-tasks agents to perform a variety of tasks [121,122]. Table 3 lists works relevant to several main ideas in ML for Reinforcement Learning (RL).

Table 3: *Reinforcement Learning (RL) Ideas, Techniques and References*

| Ideas & Techniques | Relevant Literature |
|---|---|
| Types of RL | Q-Learning[119], SARSA [118], Temporal Difference (TD)-Learning[123] |
| RL Algorithm | Self-Training [124], Deep Q-Learning (DQN) [125], Deep Deterministic Policy Gradient (DDPG) [126], Offline [127] |
| Apprenticeship Learning Efficient RL | SayCan[128], Q-Attention [129], Imitation Learning [130] Replay with Compression [131], Decision Transformer [132] |
| Evolving Curriculum | Adversarially Compounding Complexity by Editing Levels (ACCEL) [133], Paired open-ended trailblazer (POET) [134], Autonomous Driving Scene Render (READ) [135] |
| Bandit Problem | Bandit Learning [136], Batched Bandit [137], Dueling Bandit [138], Upper Confidence Bound (UCB) [139] |

Many human teaching or learning techniques have been the source of inspiration for advancement of this thrust. With the emergence of effective sampling techniques amongst others, the Efficient Zero [140] models have been able to progress by accumulating experiences through randomly playing against itself repeatedly, which improves the ability of game-playing models. Reinforcement learning has not only made breakthroughs in all-information open games; recently there has also been breakthrough in multi-player partial information games like Texas Hold'em and Alpha Hold'em [117]. For multi-agent reinforcement learning, the common benchmark task StarCraft Multi-Agent Challenge (SMAC) [141] can now be effectively completed by reinforcement learning models, which effectively decomposes the cooperation task into role-learning by large neural networks [142] amongst many other techniques [143,144]. This is a breakthrough for multi-agent reinforcement learning.

## 3.2 On Quantum Machine Learning

Quantum ML is one of the big next steps of ML [145]. While error-correction still limits our ability to build a fully quantum computer, it is possible to innovate with hybrid algorithms that uses quantum sub-algorithms or components to speed-up, robustify ML or simply to expand the theoretical boundaries of ML with 2 norm probabilities. In quantum computing, we can compute the similarity between feature vectors with state overlaps (denoted by bra and ket) instead of kernels via inner product. Consider a simple quantum ML scenario below:

$$K(x, x') = \langle \phi(x), \phi(x') \rangle \rightarrow QKE(x, x') = \langle \phi(x) | \phi(x') \rangle \tag{3.1}$$

The feature space in quantum ML can be obtained by state preparation. For instance,

$$\phi : [x_1; x_2] \rightarrow (x_1 |0\rangle + x_2 |1\rangle) \otimes (x_1 |0\rangle + x_2 |1\rangle) \tag{3.2}$$

The corresponding circuit is denoted by,

$$S_x^{AB}(|0\rangle_A \otimes |0\rangle_B) = S_x^A |0\rangle_A \otimes S_x^B |0\rangle_B \tag{3.3}$$

We can have quantum kernel estimation (see Eqn. 1) [146], or quantum feature spaces [147] in hybrid ML algorithms or intermediate scale hybrid machines [148]. This offers a new insight to the types of kernels and linear algebra that we can use to improve ML in the classical sense. Quantum physics or chemistry can be more effectively simulated in the primitive sense using quantum ML algorithm. The hybrid ML coupled with quantum material science is potentially an important stepping stone for material scientists and computer scientists alike to innovate and research more efficiently.

### 3.3 Theory, Explainable AI and Verification

In classical computer science, the very hard case of Travelling Salesman Problem (TSP), a classical NP problem, has been solved with very satisfactory result based on neural networks, which either blend with pre-training of a solver of a mini-TSP or a reinforcement learning-based [149] strategy selector combined with heuristic. Other prominent NP problems like Maximum Independent Set (MIS) or Satisfiability Modulo Test (SMT) have also been solved satisfactorily with ML-guided heuristic search [150]. This demonstrates that ML models have been able to push through boundaries that have been set forth by traditional theoretical computer science. This breakthrough has been made possible by effective latent representation learning of essential features of the problem itself and the solver.

Explainability XAI techniques like Integrated Gradients (IG) [151], Local Interpretable Model-agnostic Explanations (LIME)[152], Shapley Additive Explanations (SHAP)[153], SimplEx [154] and various others have gained much attention. LIME attempts to identify hot areas in the image responsible for features that result in the prediction. SimplEx [154] is an explainability technique that attempts to explain a prediction with linear combinations of samples drawn from the corpus of training data; the technique returns a combination of training samples that has contributed to the predictions. There are also efforts to incorporate explainability by adding a layer at the end of neural networks for capturing explainability information. Explainable Graph Neural Network NN) techniques that apply specifically to Graph Neural Networks are broadly classified into several classes: Gradients/features based Guided Back-propagation (BP) [155], Perturbation Based GNN Explainer [156], SubgraphX [157], Decomposition Based, Surrogates GraphLIME [158] and Generation [159]. These GNN XAI techniques are well-suited for explaining feature importance for predictions at either the node level, edge level or graph level.

Verification is important for protecting neural network models against adversarial behaviours; adversary behaviours can be characterized by ill-intent shifts of planes of separation in the model so that it is more likely to err on otherwise correctly classified samples or corrupting input samples with noise or otherwise. Neural network robustness verification techniques like Rectified Linear Unit-Plex (ReLUPlex) [160] and alpha-beta CROWN [161] have also made huge progress. It is a numerical bounds back-propagation technique where the score boundaries for each class are back-propagated throughout the network to determine the overlap between class scores. Specifically, in the non-linear portions of the neural network, the ReLU activation functions were bounded with linear functions. Safety-critical applications have also been secured with neural network verification techniques, and the Airborne Collision Avoidance System for Unmanned Aircraft (ACAS Xu) [162] is an ensemble of 45 neural networks whose purpose is to give anti-collision advice to flying planes, and utilize ReLUplex methods to make their advice robust.

### 3.4 Stack Optimizations for Deep Learning

Graphical Processing Units (GPU) are processors capable of parallel processing instructions. Standard GPU deep learning speedup techniques include convolutional layer reuse, featuremap reuse and filter reuse, and memory access is a common bottleneck. [163] The basic idea is that functions that are computed many times should be optimized on all levels, from high to low, including the instruction set level. The entire software stack, compiler technologies, and code generation have been optimized for deep learning computations on GPU. Deep learning GPU is known for its high energy usage; reducing energy usage is an essential objective for GPU optimization research [164]. The requirement for the scale of hardware architecture for ML is also loosening up, as engineers are packing engineering insights from large systems into smaller and energy-conserving systems, TensorFlow Lite Micro [165].

ML theory and practice have made massive progress in recent years. It is now transforming the scientific methods and has become deeply integrated with many scientific and humanities [166] fields. Application-wise, ML models have been trusted to make more and more crucial decisions for the well-functioning of society. For instance, in the criminal justice setting [167], ML models have been used to set bail for defendants; in the finance sector, models can help make decisions [168]; in the energy sector, they predict power generation efficiency for wind power stations. While neural network might still be a black box and can be hard to verify at times, its effectiveness as a predictor and sometimes generator has already been relied upon by many societal sectors for greater efficiency and effectiveness.

## 4 Development Trend of Machine Learning for Materials Science

ML has helped material scientist achieve their study aims in a wide variety of tasks, most prominently as a screening tool in the design of a large variety of materials, which include: energy materials, semiconductors, polymer design, catalyst, high entropy alloy, etc. The trend of going from processing a single dataset to achieving a specific aim to learning a latent representation of the underlying structure, which can later be finetuned to perform specific tasks, such as predicting the energetically stable structure across datasets, is rather prominent.

### 4.1 From Numerical Analysis to Feature Engineering

Traditionally, ML has been used as an advanced numerical regression tool to analyse experimental data in material science and many other fields [169, 170]. The remarkable ability of ML to interpolate data has allowed scientists to explain phenomena and verify hypotheses effectively.

Traditional material science ML practitioners often concern themselves with explicit feature engineering of specific materials [171]. Bhadeshia [171] has outlined four categories of models in material science; traditionally ML models are "models used to express data, reveal patterns, or for implementation in control algorithm". The classical works that involve material property prediction mostly fall into the fourth category. Figure 4 illustrates the feature engineering process for material science, which encompasses four stages: feature extraction; feature analysis; correlation and importance analysis; and feature selection [172].

In material space, there are many degrees of freedoms, such as the atomic coordinates, coordination numbers, interatomic distances, the position of the various species, and so on. Often, they are impractical to be used as the direct inputs to the algorithms, as they are not invariant under translation and rotation. In feature extraction (see Figure 4a), we seek to convert them into descriptors, which extract the underlying symmetry and distinguish systems that are truly different and not just a product of translations and/or rotations.

After the features are extracted, they undergo a series of analysis to fine tune and reduce the dimensionality of the descriptors space. The four commonly used methods, shown in Figure 4b, are the filter method, embedded method, wrapper method, and deep learning method. With the analysis process completed, a mapping, as illustrated in Figure 4c, which relates the importance and correlations among the selected features, can be used to visualize their dependence. In turn, this aids the process of feature selection, in which many suitable subsets of features (see Figure 4d) are chosen to proceed to the next stage - fed into the ML algorithm and compared to obtain the best performing minimal subset.

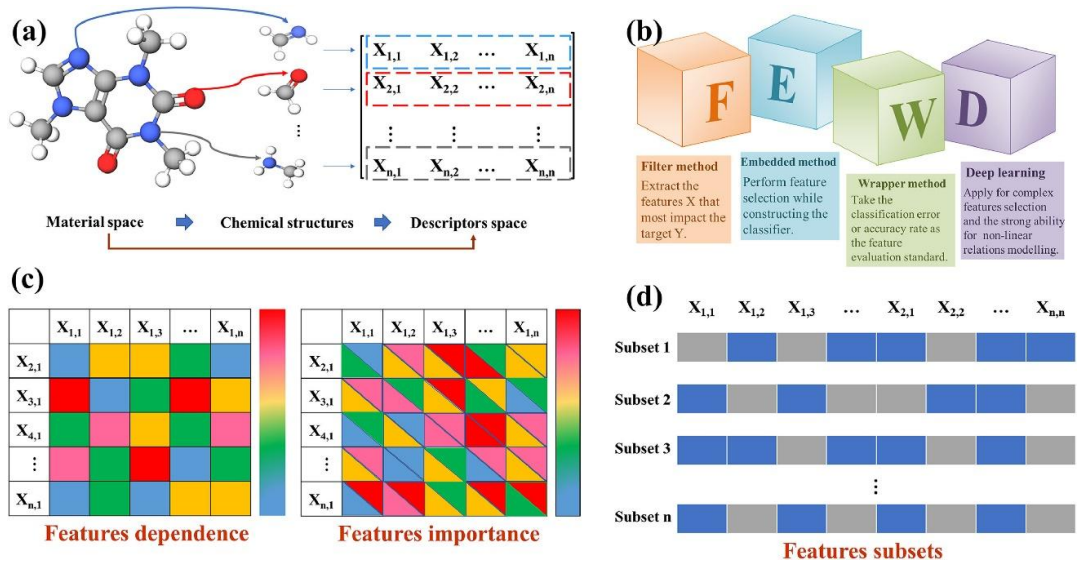

Figure 4: Feature engineering for ML applications: (a) Feature extraction process. Starting from material space, one can extract information from material space into chemical structures then to descriptors space. (b) Typical ML feature analysis methods. "FEWD" refers to Filter method, Embedded method, Wrapper method, and Deep learning. (c) Correlation and importance analysis of selected features. The feature correlations is visualized in the diagram on the left. Diagram on the right is normalized version of left diagram, where the colors indicate the relative correlation of every other feature for prediction of the row/column feature. (d) Various feature subsets obtained from feature engineering analysis. One can construct features with linearly independent combination of subsets, in other words, subsets of features are basis. Reproduced with permission [172]. Copyright 2021, Elsevier.

## 4.2 From Feature Engineering to Representation Learning

While explicit feature engineering is a practical and valuable task, it often restricts the type of task that ML can perform and does not fully use its ability to learn a generalized representation or sound separation of features and ability to interpolate or extrapolate along those dimensions. Moreover, the task of sifting through a vast dataset is laborious and hard to manage for individuals. Furthermore, with the ever-expanding computing power, the dimensionality of the features that is computationally feasible also rapidly scales up, allowing the consideration of more factors, which ultimately improves the accuracy of the prediction while also widening the coverage of material types screened. Thus, there is a push towards representation learning, an automation of feature engineering of a large material dataset [173], which better captures the internal latent features [174]. This trend encouraged a deeper integration in both development trends in ML and material science, coupled with a concise selection of ML tools, which require an intuitive understanding of mathematical and theoretical computer science ideas behind these tools.

In representation learning, the features are automatically discovered and extracted from the raw data, and thus complicated patterns that are hidden from the human user but are highly relevant could boost the accuracy and efficieency of the ML model, which is highly dependent on the quality of the selected features. Therefore, representation learning excels in applications where the data dimensionality is high and features extraction is difficult, such as speech recognition and signal processing, object recognition, and natural language processing [175].

Neural networks can be packed into layers or attention blocks that can be integrated into a single neural network. Effective embedding of information that is a dimensional reduction tool reduces the complexity of the model, when upended upon the training pipeline, brings us to end-to-end learning. Figure 5 shows a simplified pipeline for material science end-to-end model, where datasets are turned into vectors by the encoder to use as the input for the surrogate model, which attempts to identify the latent representation that can be decoded to generate predictions.

Representation learning has been applied in materials science. By using the raw experimental X-ray absorption near edge structure (XANES) spectra, Routh et al. [176] managed to obtain latent features after performing unsupervised ML methods. The raw experimental data are fed into an autoencoder that includes the encoder and decoder, which uses the input data as the output data, while information is passed through a bottleneck layer, as illustrated Figure 6.

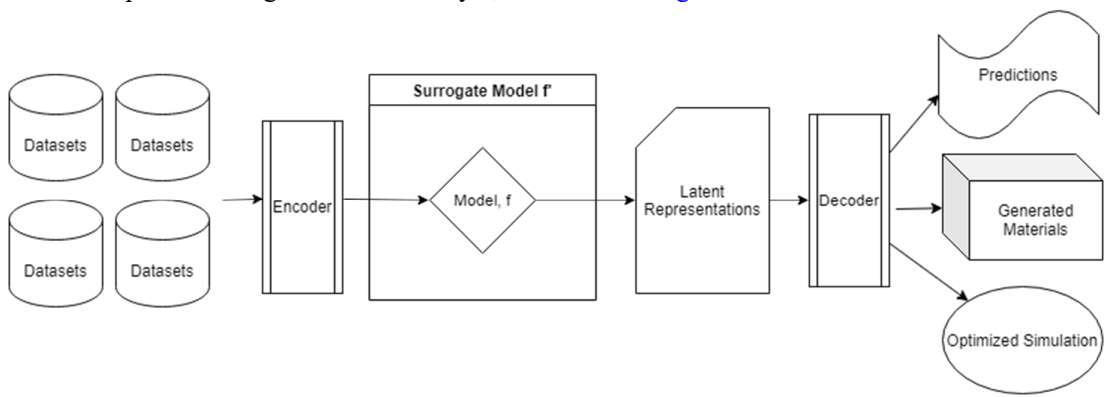

Figure 5: Infographic of End-to-End Model. End-to-End models take multi-modal dataset as inputs, and encodses them into vectors for the surrogate model. The surrogate model then learns the latent representation, which makes the internal patterns of these datasets indexable. One is then able to decode the latent representation into an output form of our choice, which includes property predictions, generated novel materials and co-pilot simulation engines.

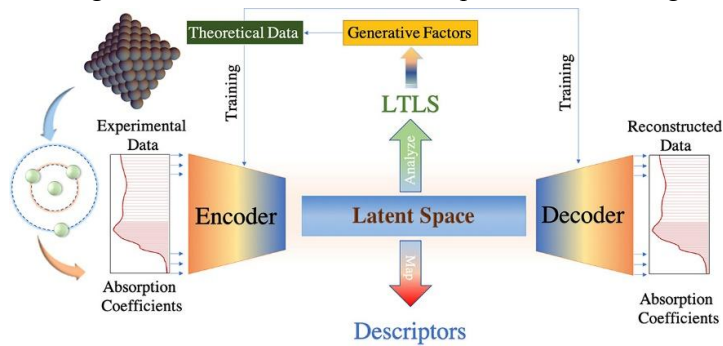

Figure 6: Schematic of the representation learning methods used in the structural characterization of catalysts, where the autoencoder, which includes the encoder and decoder, is used, with the input and output data being the same. Reproduced with permission [176]. Copyright 2021, American Chemical Society.

## 4.3  From Representation Learning to Inverse Design

After learning the representations that are critical in influencing the functionality of the materials, we ought to think: could we use them inversely, to generate novel and maybe better materials? This question had been sought in 1999 by Franceschetti and Zunger [177], where they successfully searched for the alloy of fixed elements with targeted electronic structure, using monte carlo method only. This limited yet profound results show us the vast usefulness of solving the inverse problem. Now, armed with the computation power and advancement in machine learning, we are in a better position to answer this question. Generative models like variational autoencoder (VAE) and generative adversarial network (GAN) have been applied in the inverse design of molecules and solid-state crystal.

By combining the power of representation learning and generative models into a single extensive model, that is the joining of neural networks from several parts of the workflow into a single network, many benefits can be reaped. First of all, the ability of an extensive network to counter noise levels in the training dataset, resulting in better predictions or better-generated solutions. Secondly, the latent representation learnt from each part of the pipeline is more consistent with the final goal of experimentation or design. Thirdly, the absence of human error-prone non-ML intervention helps experimenters focus on the overall goal and architecture.

By using discriminative models, generative models, and rapid simulation, whether standalone or in combination, one can construct sophisticated models that tackle problems ranging from predicting density functional theory (DFT) properties to inverse device design with confidence. One can also explore material design at different scale and granularity with ML model as an aide. An example of this is shown in Figure 7, where both discriminative and generative models are used jointly to design photonics. When the dimensionality of the photonic structures involved is very low, at the order of 1, analytical methods are well-suited. However, as the dimensionality increases, the analytical methods are no longer feasible, and the ML methods are required. On its own, discriminative models are suitable at slightly larger parameters space, but when the degree of freedom scales up considerably, generative model can be employed to reduce the dimensionality.

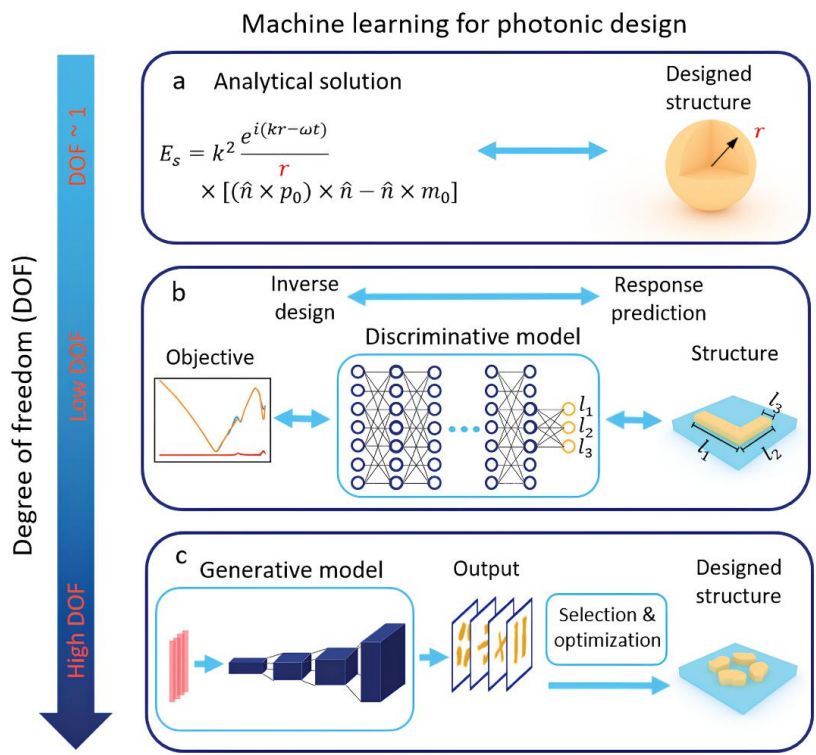

Figure 7: Depending on the degree of freedom (DOF) involved, the machine learning methodologies of the photonic design vary. The analytical methods that are suitable for DOF of order unity are replaced by the discriminative model of ML. As DOF increases, generative model is leveraged to bring down the dimensionality. Reproduced with permission [178]. Copyright 2021, Wiley-VCH.

## 5   Databases in Material Science

Data is prevalent in material science; data which originates from every aspect and process of material science research endeavour have varying types, accuracy and genre. Table 6 lists the typical data types and database that are used in ML models. A material science task often includes processing a combination of data types listed.

The broad spectrum of data types and multi-modules of input data dictates that material science models need to learn to integrate multi-modal data to produce meaningful research results. This trend also means that the material science community needs to embrace the software and statistical revolution that will propel the field forward.

Table 6: Typical Material Science Databases

| Data Type | Database |
|---|---|
| Computational Data | OQMD: Materials Properties Calculated from DFT [179,180], Materials Project [181], Joint Automated Repository for Various Integrated Simulations (JARVIS) [182], AFLOW[183], MatCloud[184], MPDS[185], NOMAD[186], C2DB[187], 2DMatPedia [188] |
| Crystallographic Data | ICSD [189], Crystallography Open Database (COD) [190], The NIST Surface Structure Database (SSD4) [41], Aspherical Electron Scattering Factors [191], AlphaFold [192] |
| Imaging/Spectra Data | MatBench[193], TEMImageNet [194], Single-Atom Library [195] |
| Other types | Knowledge Graph e.g. propnet[196] |

In order to use computer systems to process material information, material-related nomenclatures have to adapt to computer processing norms, like string. Both atomic and structure of molecules should be evident by parsing strings, e.g. Simplified molecular-input line-entry (SMILES), BigSMILES [197], Self-referencing embedded strings (SELFIES) [198], Physical Information File (PIF) [199]. Material Science datasets are often implemented in neural network data loaders like Deep Graph Library [200]. The ML community's Datasets have codebases that organize information that eases software engineers to call and process with a library. Most quantum chemistry software is softwareengineering based Application Programming Interface (API) to share and process Quantum Chemistrydata; it is written to store quantum calculation data with well-tested and scalable database norms (like schema) and eases or speedup data batch processing. The basis for quantum chemistry libraries is standardized; typical ones include Gaussian Orbital Basis (GTO), Plane Wave Basis (PW), and Numerically Tabulated Atom-centered Orbitals (NAO). Table 7 lists softwares which might be useful. The first portion lists general deep learning libraries (APIs), second portion lists useful libraries for machine learning tasks, third portion lists tools that might be useful to material science.

Table 7: Machine Learning Libraries. All descriptions were adapted from the references therein.

| Library | Library | Description |
|---------|---------|-------------|
| General deep learning libraries (APIs) | Deepmind Jax [201] | Open ML codebase by Deepmind. With its updated version of Autograd, JAX can automatically differentiate native Python and NumPy code. |
| | Keras [202] | Free open source Python library for developing and evaluating deep learning models |
| | PyTorch [203] | PyTorch is an open source machine learning framework based on the Torch library |
| | TensorFlow [204] | Created by the Google Brain team, TensorFlow is an open source library for numerical computation and large-scale machine learning. |
| Useful libraries for machine learning tasks | HuggingFace [205] | Open NLP Library with Trained Models, API and Dataset Loaders. |
| | OpenRefine [206] | OpenRefine is an open-source desktop application for data cleanup and transformation to other formats, an activity commonly known as data wrangling. |
| | PyTorch Geometric [207] | PyG (PyTorch Geometric) is a library built upon PyTorch to easily write and train Graph Neural Networks (GNNs) for a wide range of applications related to structured data. |
| | PyTorch Lightning [208] | PyTorch Lightning is the deep learning framework for professional AI researchers and machine learning engineers who need maximal flexibility without sacrificing performance at scale. |
| | VectorFlow [209] | Optimized for Sparse Data in Single Machine Environment |
| | Weights & Biases [210] | W&B for experiment tracking, dataset versioning, and collaborating on ML projects. |
| Tools that might be useful to material science | Dscribe [211] | Provides popular feature transformations ("descriptors") for atomistic materials simulations, including Coulomb matrix, Ewald sum matrix, sine matrix, Many-body Tensor Representation (MBTR), Atom-centered Symmetry FunSction (ACSF) and Smooth Overlap of Atomic Positions (SOAP). |
| | Open Graph Database [212] | The Open Graph Benchmark (OGB) is a collection of realistic, large-scale, and diverse benchmark datasets for machine learning on graphs. OGB datasets are automatically downloaded, processed, and split using the OGB Data Loader. |
| | RDKit [213] | Opensource library for Converting Molecules to SMILES string |
| | Spektral [214] | Spektral is a Python library for graph deep learning, based on the Keras API and TensorFlow 2 |

# 6 Machine Learning Descriptors for Material Science

The material science datasets are often comprised of atomistic information with the coordinates of atoms, the charges on the atoms, and their compositions. To capture the spatially invariant information, the local environment on atomic scale is oftenly extracted, such as the list of neighbouring atoms and their relative spatial positions. They are then compactified and propagated as descriptors in the form of a vector, in a neural network which maps this information to their properties that are of interests: total energy, mass density, bulk moduli, etc. In general, a good descriptor needs to have the following qualities:

i)    Invariant under spatial transformation (arbitrary translations, rotations, and reflections)

ii)   Invariant under permutation/exchange of atoms of identical species, i.e., only a unique representation for each arrangement of atoms.

iii)  Computationally cheap and easy to implement.

iv)   Minor deviation under small perturbations in the atomic structure.

Clearly, the cartesian coordinates of the atoms do not satisfy the points i) and ii), even though it is the easiest imaginable method. There are many different descriptors that have been tried and tested in material science, which we will attempt to briefly summarize in this section, but it is by no means exhaustive. For further information and use examples on descriptors, the reader is recommended to the articles of Li et al. [215] and Schmidt et al. [216].

## 6.1 Pair-wise Descriptor

Pair-wise descriptor is a type of descriptor that considers each and every possible pair of atoms in the system. Examples include Z-matrices, Weyl matrices, and more recently, the Coulomb matrices [216]. A figure briefly describing the Weyl matrices and Coulomb matrices are shown in Fig. 8(a) below. In the work of Rupp et al. [217], Coulomb matrices were constructed for a set of organic molecules that are numerically extracted and sorted descendingly, then the Euclidean difference between the vectors of eigenvalues are computed and defined as the distance between two molecules (with different dimensions accounted for by adding trailing zeroes to vectors). Using this as the sole descriptor, they developed a ML approach for fast and accurate prediction of molecular atomization energy. The same eigenvalue-based method has also been used in a number of recent studies [218, 219]. The downsides of this method are the inability to differentiate enantiometer [220] and the loss of information, as the dimensions are reduced from N^2 to N, which can sometimes be an advantage [219].

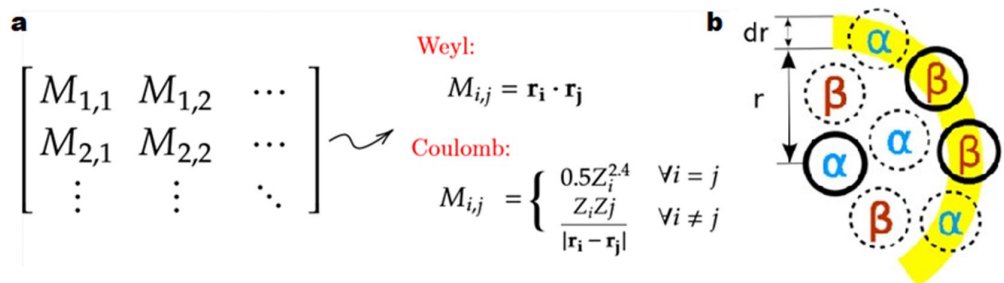

Figure 8: (a) The mathematical description of the Weyl and Coulomb matrices. (b) The construction of the PRDF sums, where atoms covered by the yellow strip covering the radius $(r, r + dr)$ are considered [221]. Figure (b) is reproduced with permission [221]. Copyright 2022, American Physical Society.

As described, the Coulomb matrices methods are only viable for finite system. To extend to the pairwise descriptor to infinite periodic system, Faber et al. [220] proposed three different methods: Ewald sum matrices, Sine matrices, and Extended coulomb-like matrix, and their results show that Sine matrix is the most efficient and outputs the smallest error.

Another alternative, the partial radial distribution function (PRDF) was proposed by Schütt et al. [221] and used in their work to perform fast prediction of density of states at Fermi level for different types of solids. The pairwise distances $d_{\alpha\beta}$ between two atoms type are considered, in the following equation for PRDF:

$$g_{\alpha\beta}(r) = \frac{1}{N_\alpha V_r} \sum_{i=1}^{N_\alpha} \sum_{j=1}^{N_\beta} \theta\left(d_{\alpha_i\beta_j} - r\right) \theta\left(r + dr - d_{\alpha_i\beta_j}\right) \tag{6.1}$$

where $\theta(x)$ is the step function, $V_r$ is the volume of the primitive cell, while $N_\alpha$ and $N_\beta$ are the number of atoms of types $\alpha$ and $\beta$. Only the atoms in the primitive cell are considered as the shell center, i.e., the atoms $\alpha_i$, see Figure 8(b). This function is invariant under translation, rotation, and different choice of the unit cell.

## 6.2 Local Descriptor

The most intuitive methods to describe a system of atoms that also take into the geometrical aspect into account is the neighbour-based or local descriptor, as the electron density is only weakly affected by distant atoms. By considering the neighbouring atoms of a selected atom within a pre-determined cutoff radius, we can store the information about their bonds, such as the bond distance and angle.

Behler and Parinello [222] proposed the use of two symmetry functions, the radial symmetry function $G_i^1$ and angular symmetry function $G_i^2$:

$$G_i^1 = \sum_{j \neq i}^{all} e^{-\eta(R_{ij}-R_s)^2} f_c(R_{ij}) \tag{6.2}$$

$$G_i^2 = 2^{1-\zeta} \sum_{j,k \neq i}^{all} \left(1 + \lambda \cos\theta_{ijk}\right)^\zeta \times e^{-\eta\left(R_{ij}^2 + R_{ik}^2 + R_{jk}^2\right)} f_c(R_{ij}) f_c(R_{ik}) f_c(R_{jk}) \tag{6.3}$$

where $R_{ij}$ is the distance between atom $i$ and $j$, $\theta_{ijk}$ is the angle between the three atoms $i, j, k$. There are four free parameters in total, $\lambda (= +1, -1)$, $\eta$, $\zeta$, and the implicit $R_c$ in $f_c$, defined as

$$f_c(R_{ij}) = \begin{cases} 0.5 \times \left[\cos\left(\frac{\pi R_{ij}}{R_c}\right) + 1\right] & \text{for } R_{ij} \leq R_c \\ 0 & \text{for } R_{ij} > R_c \end{cases} \tag{6.4}$$

The symmetry functions capture the local environment of an atoms, are invariant to permutation, translation, rotation, and changes in coordination number. They have been used in reproducing potential energy surface (PES) at DFT accuracy. This formalism was extended and studied in extensive details in Behler [223], where the set of symmetry functions are coined the "Atom-centered Symmetry Functions (ACSFs)". A further generalization was done by Seko et al. [224], which included basis functions other than the Gaussian in Equation 2, such as Neumann functions and Bessel functions. They also introduced the use of the Least Absolute Shrinkage and Selection Operator (LASSO) technique to optimize the basis set and find the sparsest representation to speed up computation. This was successfully used to reproduce almost DFT-accuracy phonon dispersion and specific heat for hcp Mg. A more recent work to reduce the undesirable scaling in ASCF has also been discussed [225].

Another approach using bispectrum, a three-point correlation function, was introduced by Bartók et al. [226]. In this approach, they first construct local atomic density function for each atom i, as

$$\rho_i(\boldsymbol{r}) = \delta(\boldsymbol{r}) + \sum_j \delta(\boldsymbol{r} - \boldsymbol{r}_{ij}) f_c(|\boldsymbol{r}_{ij}|) \tag{6.5}$$

where the $\delta(\boldsymbol{r})$'s are the Dirac Delta function. This atomic density is then projected onto the surface of a 4D sphere, by expanding the atomic density using 4D spherical harmonics, $U_{m'm}^j$ (index $i$ omitted):

$$c^{j'}_{m'm} = \left\langle U_{m'm}^j \middle| \rho \right\rangle \tag{6.6}$$

and the bispectrum is then built from these coefficients, defined as:

$$B_{j_1, j_2, j_3} = \sum_{m'_1, m_1 = -j_1}^{j_1} \sum_{m'_2, m_2 = -j_2}^{j_2} \sum_{m', m = -j}^{j_1} \left(c^j_{m'm}\right)^* C^{jm}_{j_1 m_1 j_2 m_2} C^{jm'}_{j_1 m'_1 j_2 m'_2} c^{j_1}_{m'_1 m_1} c^{j_2}_{m'_2 m_2} \tag{6.7}$$

where the $C^{jm}_{j_1 m_1 j_2 m_2}$'s are the ordinary Clebsch-Gordan coefficients of the SO(4) group.

The Smooth Overlap of Atomic Positions (SOAP) descriptor [226] uses the atomic density defined in Equation 5, but with the Dirac Delta function replaced by the Gaussians, expanded in terms of spherical harmonics:

$$\exp(-\alpha|\boldsymbol{r} - \boldsymbol{r}_i|^2) = 4\pi \exp\left[-\alpha(r^2 + r_i^2)\right] \sum_{lm} h_l(2\alpha r r_i) Y_{lm}(\hat{\boldsymbol{r}}) Y_{lm}^*(\hat{\boldsymbol{r}}_i) \tag{6.8}$$

where $h_l$'s are the modified spherical Bessel functions of the first kind and $Y_{lm}$ is the spherical harmonics. A similarity kernel $k(\rho, \rho') \equiv \int d\hat{R} \left| \int \rho(\boldsymbol{r}) \rho'(\hat{R}\boldsymbol{r}) d\boldsymbol{r} \right|^n$ was introduced to compare two different atomic environments, where $n = 2$ is used in their study. The normalized kernel or SOAP kernel

$$K(\rho, \rho') = \left( \frac{k(\rho, \rho')}{\sqrt{k(\rho, \rho) k(\rho', \rho')}} \right)^\xi \tag{6.9}$$

where $\xi$ is any positive integer, chosen to control the sensitivity, goes into the PES of the form

$$\varepsilon(\boldsymbol{q}) = \sum_{k=1}^N \alpha_k K(\boldsymbol{q}, \boldsymbol{q}^{(k)}) \tag{6.10}$$

where the $\boldsymbol{q}^{(k)}$ is the training set configurations. The SOAP descriptor is now widely adopted, especially in the machine-learning of potentials [227-230].

Based on the SOAP approach, Artrith et al. introduced another descriptor for machine-learnt potentials, which does not scale with the number of chemical species, a feature that SOAP lacks [216]. This is carried out by taking the union of a set of invariant coordinates which maps the atomic structure and another one that maps the chemical composition, which are both described by the radial and angular distribution functions:

$$\mathrm{RDF}_i(r) = \sum_\alpha c_a^{(2)} \phi_\alpha(r), 0 \le r \le R_c \tag{6.11}$$

$$\mathrm{ADF}_i(r) = \sum_\alpha c_a^{(3)} \phi_\alpha(\theta), 0 \le \theta \le \pi \tag{6.12}$$

where $R_c$ is the cutoff radius and the $\phi_\alpha$ is a complete basis set, which in their work is the Chebyshev polynomials of the first kind.

### 6.3 Graph-based Descriptor

By converting the atoms and bonds in a molecule into vertices and edges, we can turn the molecule into a graph as depicted in Figure 9a. The information about the edges and the edge distance between vertices can then be encoded into the adjacency and distance matrices [231], shown in Figure 9b. This graph-theoretic approach is known as structure graph, which has been devised long ago in 1863. Despite the simplicity and apparent loss of 3D information, structure graphs have seen widespread uses in comparing the structure of molecules.

The generalization of structure graph to periodic systems is the Universal Fragment Descriptor (UFD) [232], which uses the Voronoi tessellation (see Figure 9c) to determine the connectivity of atoms, in the following 2 steps:

i) The crystal is partitioned into atom-centered Vornoi-Dirichlet polyhedral.

ii) Atoms that share a perpendicular-bisecting Voronoi face with interatomic distance smaller than the Cordero covalent radii (with 0.25 Å tolerance) is determined to be connected. Periodic atoms are considered.

which defines the graph. Subgraphs are also constructed corresponding to the individual fragments, which include linear paths connecting at most 4 atoms and circular fragments, representing the coordination polyhedral of an atom. Then, an adjacency matrix $\mathbf{A}$ is constructed based on the determined connectivity, along with a reciprocal distance matrix $\mathbf{D}$ $(D_{ij} = 1/r_{ij}^2)$, which when multiplied together gives the Galvez matrix $\mathbf{M} \equiv \mathbf{A} \cdot \mathbf{D}$. The information about the atomic/elemental reference property $\mathbf{q}$ (could be Mendeleev group and period number, number of valence electron, electronic affinity, covalent radii, etc) is then incorporated in the pair of descriptors for a particular property $\mathbf{q}$:

$$T^E = \sum_{i=1}^{n-1} \sum_{j=i+1}^{n} |q_i - q_j| M_{ij} \qquad (6.13)$$

$$T^E_{bond} = \sum_{\{i,j\}\in\text{bonds}} |q_i - q_j| M_{ij} \qquad (6.14)$$

where the former runs over all pairs of atoms while the latter only considers bonded pairs of atoms.

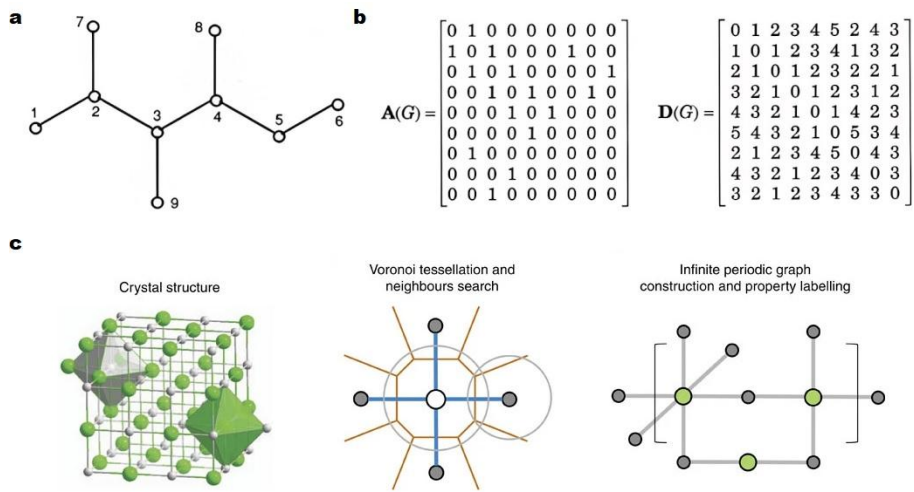

Figure 9: a) Structure graph for 2,3,4-trimethylhexane and (b) the related adjacency and distance matrix. Adapted with permission [231]. Copyright 1992, American Physical Society. (c) The Universal fragment descriptors. The crystal structure is analysed for atomic neighbours via Voronoi tessellation with the infinite periodicity taken into account. Reproduced with permission [232]. Copyright 2017, Springer Nature.

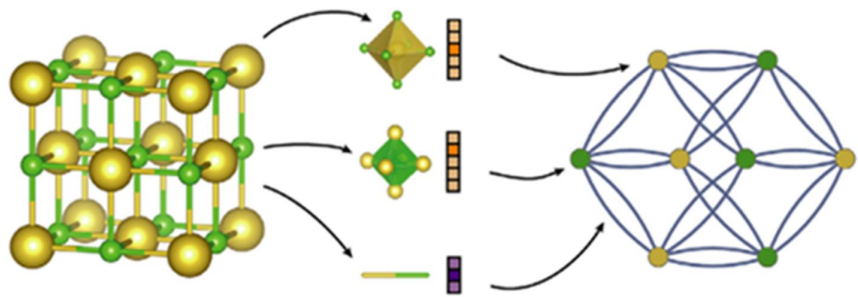

Figure 10: Crystal graph construction proposed used in the generalized crystal graph convolutional neural networks. Reproduced with permission [233]. Copyright 2018, American Physical Society.

Xie et al. proposed a framework, the generalized crystal graph convolutional neural networks (CGCNN) [233] which introduced another graph-based descriptor that is inspired by the UFD. Their construction of the crystal graph is illustrated in Figure 10, where the connectivity determination is the same as in UFD, but they used the one-hot encoding to encode the atom and bond properties in two separate feature vectors: node feature vectors and edge feature vectors. They are the descriptors, which are then sent through convolutional layer, which further extracts critical features while reducing the dimensions. Convolutional neural network is discussed in Section 7.

### 6.4 Topological Descriptor

Topology famously does not differentiate between a donut and a coffee mug, as they both have a hole. This is because in topology, most of the geometrical features are stripped away, leaving only quantities that are invariant under continuous deformation. This seemingly bizarre concept has had deep impacts in physics and was the main theme of the works that won the 2014 Nobel prize. A branch of topology, persistence homology, measures the topological features which persist across different scales or granularity, and encodes them into diagrams. This idea has already been used to classify and describe proteins [234, 235], and used as ML descriptor for crystalline [236] and armophous solids [237]. However, it has not been widely used due to its mathematically complicated nature, and the lack of physical and chemical intuition also further hinders the ability to interpret the results [216]. Here, we will attempt to provide a simplified and non-rigorous overview of persistent homology and the crystal topological descriptor of Jiang et al. [236]. For a rigorous and detailed introduction to persistent homology, the ref. [238] is recommended along with other works cited in this paragraph, especially ref. [235].

The basic building blocks of persistent homology are the *simplices* (see Figure 11): a 0-simplex is a point, a 1-simplex is two connected points, a 2-simplex is a filled triangle, and a 3-simplex is a filled tetrahedron. A *face* can be a point, a line, or a 2D surface, depending on the number of points. A *simplicial complex K* is a collection of simplices which satisfies two conditions:

i)    Faces of a simplex in $K$ are also in $K$.

ii)   Any intersection of two simplices in $K$ is a face of both the simplices.

In a simplicial complex, holes are considered as voids that are bounded by simplices of different dimensions. In dimension 0, a connected component is counted as a hole; in dimension 1, a hole is a loop bounded by 1-simplices or edges; in dimension 2, hole is bounded by 2-simplices or triangles. The number of $i$-dimensional holes or voids in a simplicial complex is basically described by the $i$-th Betti numbers, $\beta_i(K)$, e.g., $\beta_0$ is the number of connected components, $\beta_1$ is the number of loops, and $\beta_2$ is the number of cavities. An example is shown in Figure 11b, where there are five 0-simplices or vertices, forming six 1-simplices, and one 2-simplices. Since all the points are connected, $\beta_0 = 1$; there is a square-shaped hole enclosed by the 1-simplices A, B, C, and F, giving $\beta_1 = 1$. Note that the face of the triangle is filled, thus it is not counted as a hole.

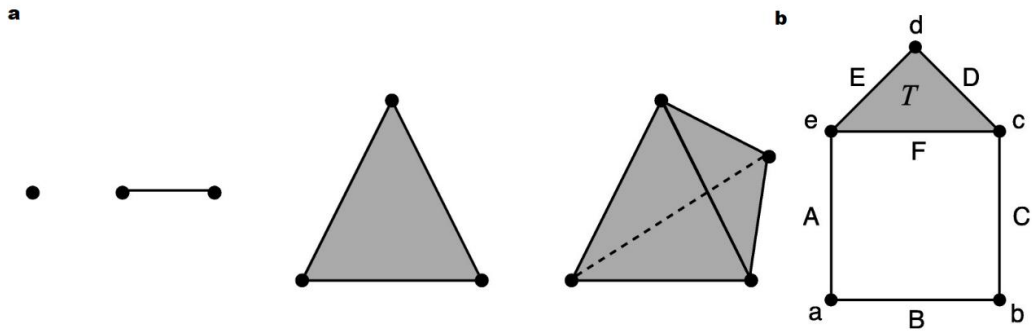

Figure 11: (a, Left to Right) 0-, 1-, 2-, 3-simplex. (b) An example of a simplicial complex, with five vertices: a, b, c, d, and e, six 1-simplices: A, B, C, D, E, and F, and one 2-simplex T. The Betti numbers for this complex are $\beta_0 = \beta_1 = 1$. Reproduced with permission [238]. Copyright 2019, Springer Nature.

To generate simplicial complex from the crystal data, we use the Vietoris-Rips (VR) filtration process, giving VR complex. This is carried out by increasing a filtration parameter, commonly the Euclidean distance cutoff between points, where points that are within cutoff distance of each other are connected. The filtration parameter used in the crystal descriptor is the radius measured from each atom, $d$, which is increased from 0 Å to 8 Å. As $d$ increases, the simplicial complex also undergoes changes, where the Betti numbers of "holes" change. This can be quantitatively plotted using persistence barcodes for each of the Betti numbers, as can be seen in Figure 12, where each barcode represents each of the "hole" for each Betti number. As $d$ reaches 4 Å, all the Betti 0 barcodes except one suddenly terminate, indicating that the points are now all connected as the Na atoms are separated by 4 Å. There is no Betti 1 barcode because the distances between any two Na atoms are the same, reflecting the structure symmetry.

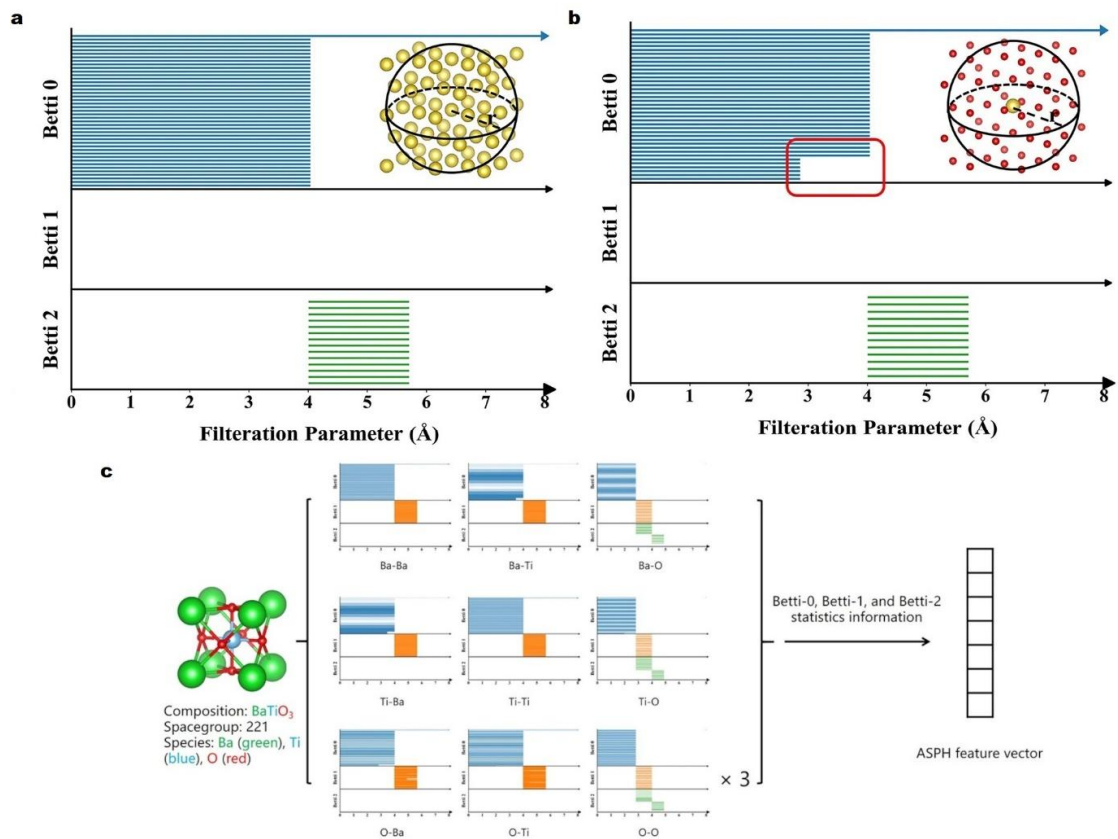

Figure 12: Persistence barcode plot for the selected Na atom inside a NaCl crystal, surrounded by only (a) Na atoms and (b) Cl atoms. (c) Construction of crystal topological descriptor, taking into account different chemical environment. Reproduced with permission [236]. Copyright 2021, Springer Nature.

To embed different elemental compositions, atom-wise chemical information is used, where a chosen atom is surrounded by atoms of a chosen type, such as in Figure 12(a) and Figure12(b), where the selected Na atom is surrounded by only Na atoms and Cl atoms, respectively. The birth, death, and the persistent length of the Betti barcodes are then encoded in a vector known as ASPH feature vector.

## 6.5 Reciprocal Space-Based Descriptor

The reciprocal space is linked to the real space by the means of Fourier transform and can be mapped using x-ray diffraction (XRD) into 2D diffraction pattern (see Figure 13a), either experimentally or computationally. 2D XRD data has first been applied as descriptor by Ziletti et al. [239] to automatically classify crystal structures. In their work, they rotated the crystal 45° clockwise and counterclockwise about a chosen axis and superimposed the obtained XRD patterns. This is then carried out for the other two axes, with different colours of the RGB palette chosen for the patterns obtained from the rotation of different axis (e.g., red for x-axis, green for y-axis, and blue for z-axis, see Fig. 13b). The final obtained pattern is then used as the descriptor, fed into a convolutional network, similar to image-based object recognition. The benefits of the XRD descriptor are that the dimension is independent of the system size and very robust to defects (compare Fig.13b and Fig.13c).

The more conventional XRD is the 1D XRD, shown in Fig. 13d for different crystals, which is obtained based on Bragg's law, mapping the 3D structures into 1D fingerprints. 1D XRD based descriptor has been used to classify crystal structure [240] and predict their properties [241]. In the latter, the group used modified XRD, where only the anions sublattice is considerd with the cations removed, and the *pymatgen* package is used to generate the XRD computationally. They successfully distinguished solid-state lithium-ion conductors with this descriptor using unsupervised learning.

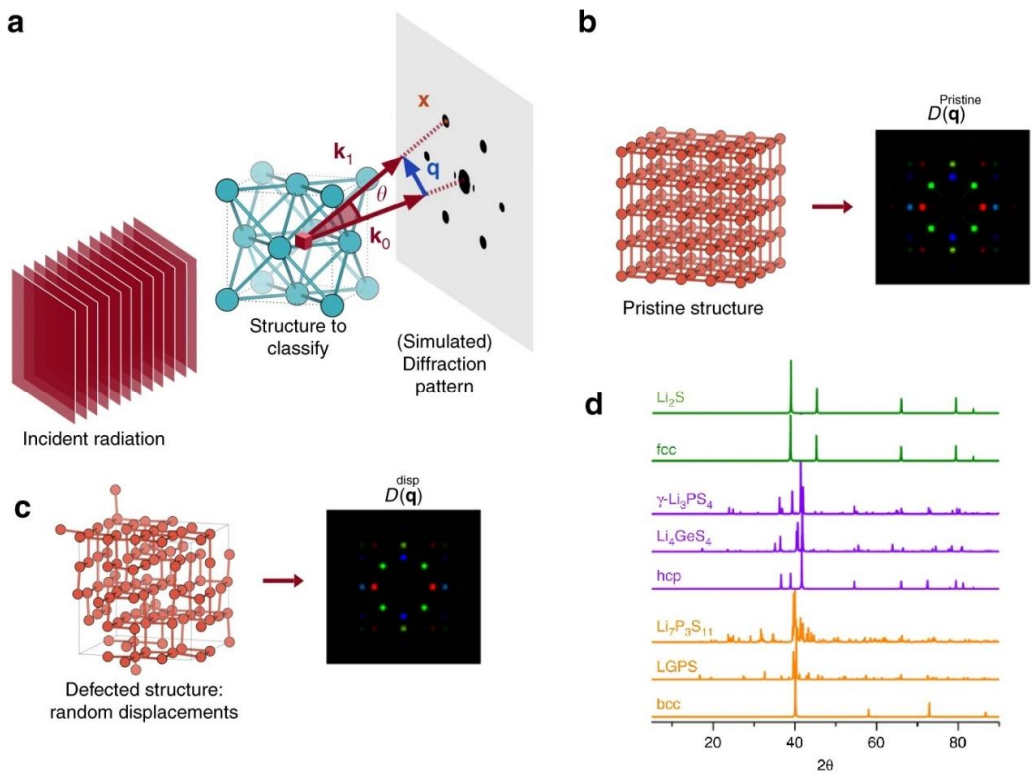

Figure 13: (a) Experimental XRD method, where X-Ray plane wave incidents on a crystal, resulting in diffraction fingerprints. (b) XRD-based image descriptor for a crystal where each RGB colour corresponds to rotation about the x, y, z axes. The robustness of the descriptor against defects can be observed by comparing (b) to (c). (d) Examples of 1D XRD. Figure (a)-(c) are reproduced with permission [239]. Copyright 2018, Springer Nature. Figure (d) is reproduced with permission [241]. Copyright 2017, International Union of Crystallography.

### 6.6 Reduction of Descriptor Dimension

In materials science, there are many possible combinations of various properties that can be used as descriptors. It is often difficult to select and fine-tune the descriptor space manually. This is a common problem in the field of ML, and several methods have been developed to tackle this issue: principal component analysis (PCA) [242, 243], least absolute shrinkage and selection operator (LASSO) [244], and sure independence screening and sparsifying operator (SISSO) [245]. However, they mainly work for models that are linear, hence not directly applicable for neural network- based models [239].

## 7 Machine Learning Algorithms for Material Science

In this section, we collate the recently developed ML-based tools and frameworks in materials science, grouping them together by the ML algorithms used. We then briefly describe the commonly used algorithms and also introduces some of the emerging algorithms, which could further unlock the potential of materials science ML applications.

## 7.1 Currently Utilized Algorithms

Table 8 enumerates the ML algorithms used in relatively recent developed tools or framework in materials science. It can be seen that the convolutional and graph neural networks are the popular algorithms, with transfer learning also picking up pace. We will briefly introduce the algorithms, drawing examples from the materials science implementation.

### 7.1.1 Kernel-Based Linear Algorithms

Support vector machine (SVM) and kernel ridge regression (KRR) are kernel-based ML algorithms, which utilize kernel functions $K(x, x')$ that allow high-dimensional feature to be used implicitly, without actually computing the feature coordinates explicitly, hence speeding up computation. Furthermore, it allows non-linear problem to be solved using linear algorithms by mapping the problem into a higher-dimensional space. Examples of commonly used kernel functions include

Linear Kernel:

$$K(x, x') = (x_i \cdot x_j + \theta) \tag{7.1}$$

Polynomial Kernel:

$$K(x, x') = (x_i \cdot x_j + \theta)^d \tag{7.2}$$

Gaussian Kernel/Radial Basis Function (RBF):

$$K(x, x') = \exp\left(\frac{-\|x_i - x_j\|^2}{\sigma^2}\right) \tag{7.3}$$

where $x$ is the input data, while $\theta$ and $\sigma$ are adjustable parameters. SVM is used for both classification and regression problem, denoted as SVC and SVR, respectively. On the other hand, KKR is used only for regression problems and it is very similar to SVR, except for the different loss functions.

Applications of both types of SVM are demonstrated in the work of Lu et al. [246]. Using atomic parameters such as electronegativity, atomic radius, atomic mass, valence, and functions of these parameters, they constructed classifier for the formability of perovskite structure, and regression models to predict the band gaps of binary compounds, using the polynomial kernel (Eq. 7.2) with $d = 2$ and $\theta = 1$.

Wu et al. [247] used KRR to assist in non-adiabatic molecular dynamics (NA-MD) simulations, particularly in the prediction of excitation energy and interpolate nonadiabatic coupling. KRR was chosen over neural networks because of the fewer hyperparameters KRR possessed, and KRR requires only the use of simple matrix operation to find the global minimum. By only providing a small fraction (4%) of sampled points, KKR gives a reliable estimate while saving MD computational effort of over an order of magnitude.

### 7.1.2 Neural Network

Artificial Neural Networks (ANNs, shortened to NNs) is a type of ML architecture that aims to mimic the neural structure of the brain. In NNs, there are 3 types of layers consists of interconnected nodes: input layer, hidden layer(s), and output layer, as shown in Figure 14a. The input layer receives the input raw data, which is then propagated to the hidden layer(s), where the nodes are functions of the backward-connected nodes and each connection is weighted. The function of a hidden layer node $m$ with $x$ being the node values of previous layer, takes the form:

$$h_m = \sigma\left(b + \sum_i^n \omega_i \cdot x_i\right) \tag{7.4}$$

where $\sigma(z)$ is known as the activation function, where a common choice is the sigmoid function $\sigma(z) = 1/(1 + e^{-z})$ and $b$ is bias term, and the ReLU (Rectified Linear Unit) function, simply defined as $\sigma(z) = \max(0, z)$. After the hidden layer(s), the information is then passed toward to the output layer, which is another function of the nodes of the final hidden layer. The outputs are measured against true value using a pre-defined cost function, with the simplest example for regression problem being the sum of squared error

$$J(\omega) = \frac{1}{n} \sum_{i=1}^{n} \frac{1}{2} (y_i - \hat{y}_i)^2$$

where $y_i$ and $\hat{y}_i$ are the true and predicted values respectively, and the sum is taken over the whole training set with $n$ being the size of the training set. The weights are then optimized iteratively using the backpropagation method, which is a function of the gradient of the cost function. For a detailed discussion, the book [248] is recommended.

Deep NNs are NNs with more than one hidden layer (see Figure 14b). By having more hidden layers, the model is better positioned to capture the nonlinearities in the data. However, having too many hidden layers can cause the convergence or learning to be slow and difficult, because the gradients used in backpropagation will tend to become vanishingly small. To overcome this issue, residual block has been devised [249], which introduces shortcut between layers, as shown in Figure 14c.

SpookyNet [250] is a DNN-based model built to construct force fields that explicitly include nonlocal effects. In their DNN architecture, the generalized sigmoid linear unit (SiLU) activation function is used, which is given by

$$silu(x) = \frac{\alpha x}{1 + e^{-\beta x}}$$

where both $\alpha$ and $\beta$ are learneable parameters. They noted that a smooth activation function is crucial for the prediction of potential energies, as discontinuities in the atomic forces would be introduced otherwise. They introduced a loss function that has 3 components: energy, forces, and dipole moments, which is minimized by optimizing the weights using mini-batch gradient descent. They also incorporated residual block which allowed them to use a large number of hidden layers.

Convolutional NNs (CNNs) is primarily used in image pattern recognition, and is different from deep NNs by having a few extra layers, which are the convolutional and pooling layers. The extra layers filter and convolute the data to capture crucial features in the data and also reduce the input dimension, which scales quickly with resolution in image recognition problems. The work of Ziletti et al. [239] uses CNN architecture, as depicted in Figure 15. The convolution layers capture elements that are discriminative and discard unimportant details.

Graph NNs is specifically designed for input data that are structured as graph, which contains nodes and edges, and can handle inputs of different sizes. There are several different types of graphs NNs, such as graph convolutional network (GCNNs), graph attention network, and Message Passing Neural Network.

### 7.1.3 Decision Tree and Ensembles

Decision tree is a supervised method for solving both classification and regression problems, which resembles a tree. A typical decision tree is shown in Figure 16a, where each internal node represents a feature or attribute, each branch contains a decision rule, and each leaf node is a class label or a numerical value, depending on the type of problem solved. The number of node layers a decision tree contains is known as depth, which needs to be tuned. An important metric used in measuring the performance of a decision tree in classification is the information gain, which is defined as the difference of the information entropy between the parent and child node; while for regression problem, the variance reduction is the performance evaluation metric for a decision tree. Decision tree is advantageous when it comes to interpretability, but it suffers from overfitting, especially when the tree is too deep and complex. It can also be overly-sensitive to data changes.

Random forest is an algorithm that combines multiple decision tree, with each of them trained on randomized subsets of samples, where both training data and features are chosen random with replacement in a process known as bootstrapping. The final decision is then made by aggregating the results from each decision tree and taking the majority vote for classification or the average for regression. The steps taken above are collectively known as bagging, which help ensure that the random forest algorithm is less sensitive to changes in the dataset and more robust to overfitting.

Gradient Boosting Decision Tree (GDBT) is another method that uses ensembles of decision trees but in sequence rather than in parallel. GBDT works by adding decision trees iteratively, with each one attempts to improve upon the errors of the previous tree. The final output from the trees ensembles is then taken by using weighted average of the decision trees outputs.

Random forest models are used in the work of Zheng et al. [251], which predicts the atomic environment labels from the X-ray absorption near-edge structure (XANES). Using the random forest classifier of scikit-learn package, they found that 50 trees ensemble gave the best performance, even better than other classifiers, such as CNN and SVC. On the other hand, GBDT has been used for regression in the topology-based formation energy predictor [236]. Also using the scikit-learn package, they added a tree to their model one at a time and used bootstrapping to reduce overfitting. This topology-based model is able to achieved a high accuracy in cross-validation, with mean absolute error of only 61 meV/atom, outperforming previous works that uses Voronoi tessellations and Coulomb matrix method.

### 7.1.4 Unsupervised Clustering

K-Means clustering is a popular unsupervised classification algorithm which aims to group similar data points together in $K$ different clusters. $K$ number of points that are known as cluster centroids are initialized randomly, and each data point is assigned to a cluster centroid that is closest in Euclidean distance to the data point. The centroids are then moved to a new location that is the arithmetic mean of the assigned data points. This repeats until convergence, i.e., there is no more movement among the centroids. The number $K$ determines the number of classes in the data, which can be known before hand if the dataset has clear distinction, e.g., metal vs non-metal, or can be optimized using the elbow method, which has an associated cost function $J$ that is optimized by the best choice of $K$. Despite its popularity, K-Means clustering has some limitations, such as sensitivity to outliers, dependence on the centroids position initialization, ineffective for dataset with uneven distribution, and predetermined number of clusters.

Several alternatives have been proposed which improves upon the limitations of K-Means clustering. Agglomerative hierarchical clustering (AHC), used in the work of Zhang et al. [241], is initialized by using each data point as a single cluster, then iteratively merged the clusters of the closest points until one big cluster is left. Then, a dendogram or a bottom-up hierarchical tree diagram, as show in Figure 16b, which can be cut at a desired precision, as indicated in the figure via a dashed line, where 7 groups are obtained. To verify the results, they performed spectral clustering, which splits the samples into chosen $K$ groups, based on the eigenvalues of the similarity matrix constructed from the data. This process is recursively applied bisectionally, and they obtained similar clusters as the AHC. There is also the mean-shift algorithm, utilized in [252].

### 7.1.5 Generative Models (GAN and VAE)

Generative models attempt to learn the underlying distribution of a training dataset, and use that to generate new samples that resemble the original data. Two popular types of generative models are Generative Adversarial Networks (GAN) and Variational Autoencoders (VAE). As can be seen from Figure 17, there are two different neural networks in both of the models: GAN contains a discriminator and generator network, while VAE has a decoder and encoder network. In GAN, random noise is injected into the generator network and subsequently outputs a sample that is then fed to the discriminator network, which is then classified as real or generated sample. The networks are trained together until the generated samples are able to convince the discriminator that the samples are real and not generated. On the other hand, VAE tries to learn the latent representations from the training data and generate new samples based on them using probabilistic approach.

A variant of GAN, Wasserstein GAN, has been applied in the work of Kim et al. [253], which generate Mg-Mn-O ternary materials which can potentially be used as potential photoanode materials. The overview of their GAN architecture is shown in Figure 18a, which after training, takes in random Gaussian noise vector $Z$ and encoded composition vector, and spits out new unseen crystals. The new crystals are then passed to a critic and a classifier, where the former computes the Wasserstein distance that measures the dissimilarity between the generated and true data distributions, which are used to improve the realism of the generated materials, while the latter ensure that the generated materials meet the composition condition. Using this model, they found 23 previously unknown new crystals with suitable stability and band gap.

An example of inverse design using VAE was demonstrated in the work of Noh et al. [254], where their proposed a two-step VAE-based generator is shown in Figure 18b. In the first step, the materials data is passed to a convolutional autoencoder, which contains 4 convolutional layers, outputting a compressed intermediate vector, which is then fed to a decoder that aims to maps the vector back to the input. The intermediate vector is fed into the VAE in $2^{nd}$ step to learn about the latent materials space. To generate completely novel polymorphs, the materials space around known stable structure is sampled using random Gaussian distributed vectors and the resulting latent vectors are decoded in a series of steps to output new stable structures. The model is able to recover 25 out of 31 known structures that are not included in the training, and 47 new valid compositions are discovered that have eluded genetic algorithms.

### 7.1.6   Transfer Learning

In materials science, high quality data for a specific type of materials is usually scarce, which severely impedes the applications of ML in generating high quality predictions [255]. Transfer Learning (TL) is a method that can be applied to overcome this data scarcity issue. In transfer learning, the parameters of a model that has been pre-trained on a large data set but with different task/purpose, are used to initialize training on another data-scarce task, such as the parameters of the models used for predicting formation energy is later used to train another task of predicting band gaps.

Chang et al. [256] combined pairwise transfer learning and mixture of experts (MoEs) framework in their model. In pairwise transfer learning, a model is pre-trained on a source task (task designed for the large dataset) and a subset of the pre-trained model parameters is used to produce generalizable features of an atomic structure, defined as a feature extractor. This extractor extracts a feature vector from an atomic structure, which can be used to predict a scalar property after passing thorugh a neural network. On the other hand, MoE contains multiple neural network models that specialize in different regions of the input space, known as "experts", and each of them are activated and controlled by a gating function. The outputs of the "experts" are then aggregated through an aggregation function. Using this architecture, the authors have performed many downstream, data-scarce tasks, such as predicting band gap, poisson ratio, 2D materials exfoliation energy, and experimental formation energies.

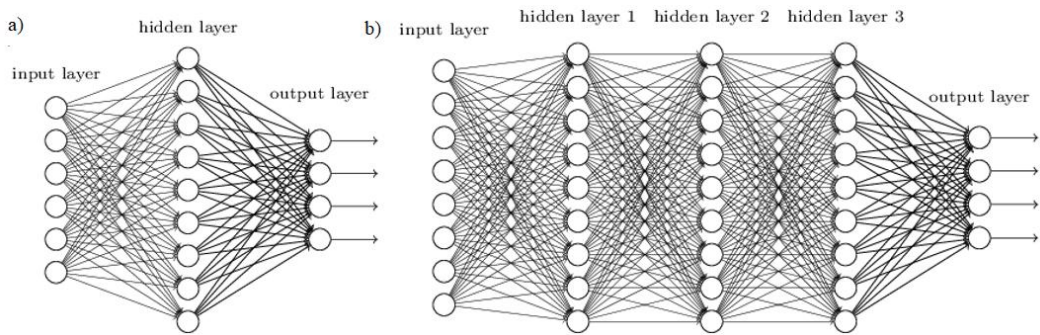

Figure 14: a) Neural network (NN) with 3 layers: input, hidden, and output. b) Deep NN with 3 hidden layers. Reproduced with permission [257]. Copyright 2015, Determination Press."

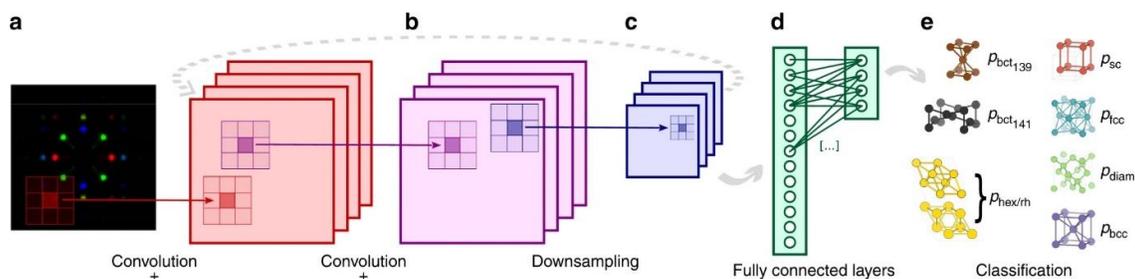

Figure 15: The CNN architecture used in the work of Ziletti et al. [239]. a) A kernel or learnable filter is applied all over the image, taking scalar product between the filter and the image data at every point, resulting in an activation map. This process is repeated in (b), which is then coarse grained in (c), reducing the dimension. The map is then transferred to regular NNs hidden layers (d) before it is used to classify the crystal structure (e). Reproduced with permission [239]. Copyright 2018, Springer Nature.

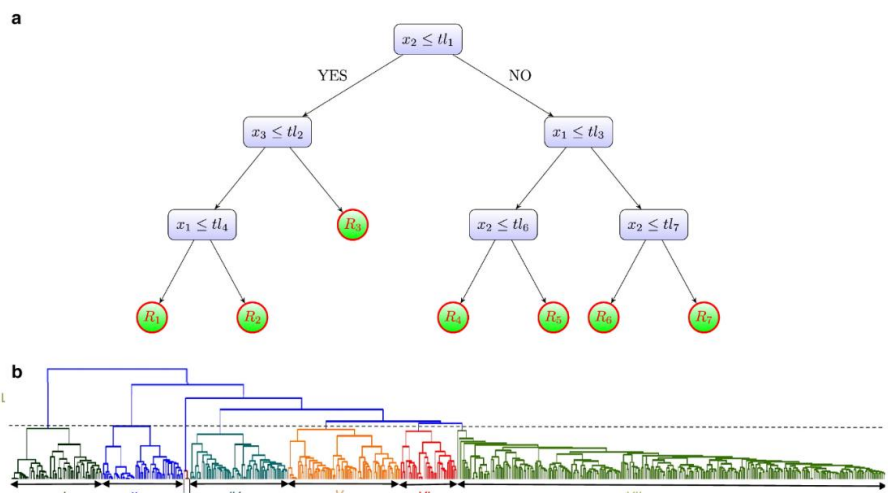

Figure 16: a) An example of a decision tree, where each square represents internal node or feature, each arrow represents branch or decision rule, and the green circles are leafs representing class labels or numerical values. b) Dendogram obtained via agglomerative hierarchical clustering (AHC) where the dashed line indicates the optimal clustering. Figure (a) is reproduced with permission [258]. Copyright 2021, Elsevier. Figure (b) is reproduced with permission [241]. Copyright 2019, Springer Nature.

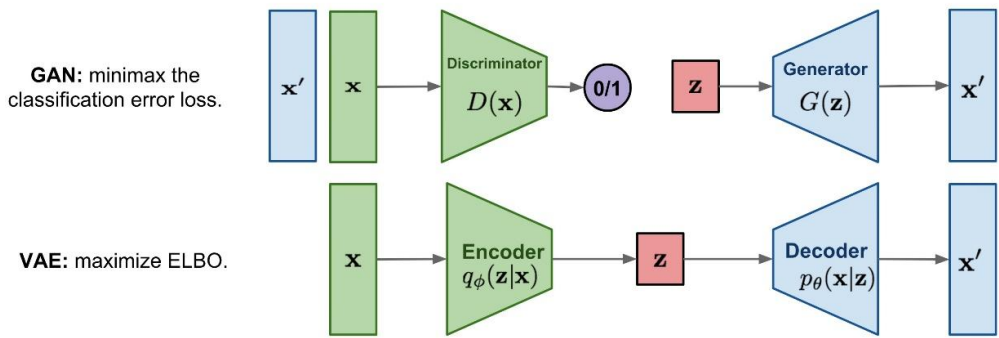

Figure 17: The architectures of the two generative models, Generative Adversarial Networks (GAN) and Variational Auto encoders (VAE). Reproduced with permission [259].

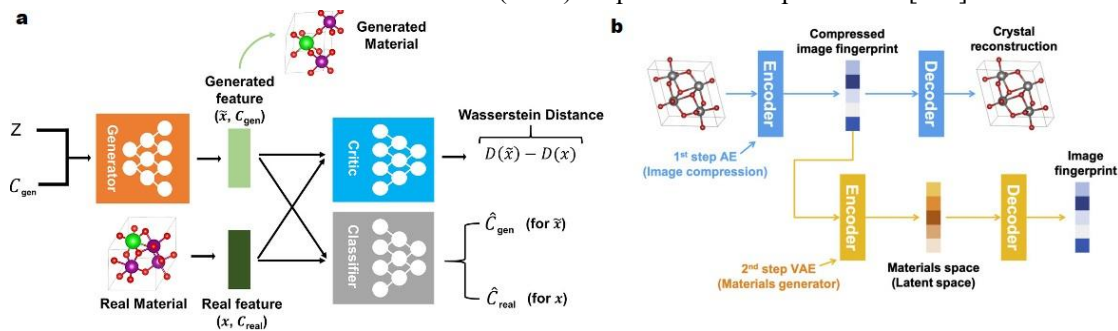

Figure 18: (a) Composition-Conditioned Crystal GAN, designed to generate crystals that can be applied in photodiode. (b) Simplified VAE architecture used in the inverse design of $V_xO_y$ materials. Figure (a) is reproduced with permission [253]. Copyright 2020, American Chemistry Society. Figure (b) is reproduced with permission [254]. Copyright 2019, Elsevier.

Table 8: List of Machine Learning (ML) algorithms used by various tools or framework developed in materials science.

| ML Algorithms | Tool |
| --- | --- |
| Support Vector Machine (SVM) | Refs. [260], [261], [262], [246] |
| Kernel Ridge Regression (KRR) | Refs. [237], [263], [247], [264] |
| Deep Neural Network | VampNet [257], DTNN [265], ElemNet [266], IrNet [267], PhysNet [268], DeepMolNet [269], SIPFENN [270], SpookyNet [250] |
| Convolutional Neural Network (CNN) | SchNet [271], Refs [239], [240], [272], [273] |
| Graph Neural Network (GNN) | CGCNN [274], MEGNet [275], GATGNN [276], OrbNet [277], DimeNET [278], ALIGNN [279], MXMNet [280], GraphVAMPNet [281], GdyNets [282], NequIP [283], PaiNN [284], CCCGN [285], [286], FFiNet [287] |
| Generative Adversarial Networks (GAN) | Refs. [288], CrystalGAN [246], MatGAN [289] |
| Variational Auto Encoder (VAE) | FTCP [290], CDVAE [291], Refs [292], [263] |
| Random Forest/ Decision Tree | Refs. [236], [293], [294], [251], [295], [296] |
| Unsupervised Clustering | Refs. [241], [282], [252], [297], [298] |

| Transfer Learning | Roost [299], AtomSets [288], XenonPy.MDL [289], |
| | TDL [290], Refs [256], [291], [292], [300], [301] |

## 7.2   Emerging ML Methods

### 7.2.1   Explainable AI (XAI) Methods

The DNNs-based approaches discussed have proved to be of great help in assisting and speeding up materials research, but their black-box nature has made understanding and explaining the results difficult, which has also plagued the general ML community [302]. In systems that trust, fairness, and moral are highly critical, such as in healthcare, finance, and autonomous driving, the decisions made by AI cannot be blindly trusted without understanding the motivation and reasoning behind the choice. Furthermore, when the black box returns results that are erroneous and puzzling, it can be difficult to diagnose and correct without knowing what exactly went wrong. To overcome these issues, the XAI techniques were introduced, which try to explain the reasonings and connections behind a prediction or classification.

There are many post-hoc (i.e., applied after model fitting) XAI methods proposed for the general ML community [303], including gradient-based attribution (Gradients, Integrated Gradients, and DeepLIFT), deconvolution-based methods (Guided Backpropagation, Deconvolution, Class Activation Maps (CAM), Grad-CAM), model-agnostic techniques (Shapley Additive explanations (SHAP), local interpretable model-agnostic explanations (LIME), Ancors).

Another type of XAI is the use of models that are inherently interpretable or explainable, which have one or more of these features [304]: sparsity, simulatability, and modularity. A model that has limited number of nonzero parameters is known as sparse, and this can be obtained by the LASSO method, whereas if a model can be easily comprehend and mentally simulate by the human user is simulatable, such as decision trees-based model. A modular model is a model that combines several modules which can be interpreted independently. In the field of materials science, the understanding of the physical and chemical intuition is paramount as it opens the door to understanding hidden connection and physics, and improve the efficiency of future studies by providing insights from previous work. The importance and implementation of XAI in materials ML tools (refer to Table 8) have been discussed in the review of Oviedo et al. [58] and Zhong et al. [302]. Zhong et al. [302] presented an overview of DNNs-based XAI as shown in Figure 19a, which highlight two fundamental motivations for XAI, which is the need for explaining how the results are obtained from the input (model processing), and what information is contained in the network (model representations). The design of an intrinsically explainable DNNs will prove important in answering the questions posed, but is itself a highly difficult task. In the following, we will illustrate some of the materials science XAI implementation, which is still in its infancy.

Kondo et al. [305] used heat maps to highlight the feature importance, particularly in identifying the positive and negative features that affect ionic conductivity in ceramics, using scanning electron microscope (SEM) images. Their CNN-based model used feature visualization method that is very similar to the deconvolution method used in CAM and Grad-CAM. By defining mask map, they obtained masked SEM images (see Figure 19b) that show features that determine low and high ionic conductivities.

A recent implementation of XAI for crystals is the CrysXPP [306] which is built upon an auto-encoder-based architure, CrysAE, that contains deep encoding module which is capable of capturing the important structural and composition information in crystal graph. The information learnt is then transferred to the GCNN contained within CrysXPP (shown in Figure 20a), which takes in feature selected from crystal graph. The feature selector contains trainable weights that selects weighted subset of important features, which is fine-tuned with LASSO to improve the sparsity of the features. An example of the explainable results obtained is shown in Figure 20b, where features that affect the band gap of GaP crystal are weighted and compared.

Compositionally restricted attention- based network (CrabNet) [307, 308] is an example of explainable DNN in materials science that is based on the Transformer-based self-attention mechanism [69], a type of algorithm initially intended for NLP, but has exploded in popularity recently. Briefly, the transformer self-attention mechanism allows the model to focus on the different parts of the input and relate them with weights to encode a representation. In this way , the dependencies between the elements are better captured, even when some of the elements are present in very small amount, e.g., dopants, which even in small quantity can have tremendous effect on the properties of the materials.

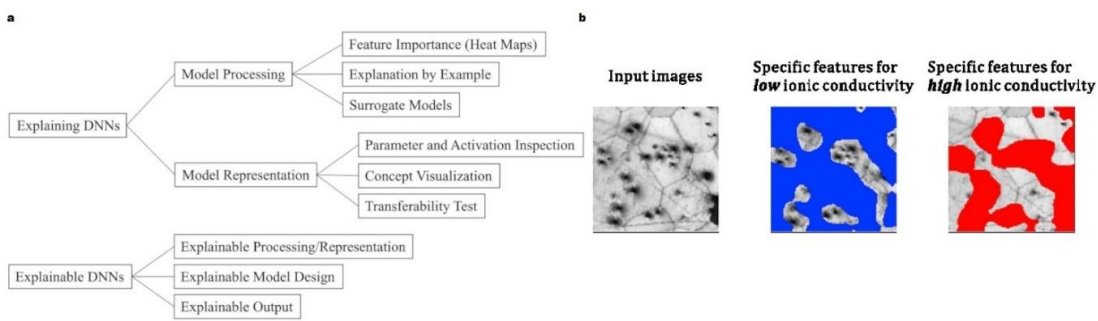

Figure 19: a) Overview of explainable DNNs approaches. b) Feature visualization in the form of heat map used in determining the ionic conductivity from SEM images. Figure (a) is reproduced with permission [302]. Copyright 2022, Springer Nature. Figure (b) is reproduced with permission [305]. Copyright 2017, Elsevier.

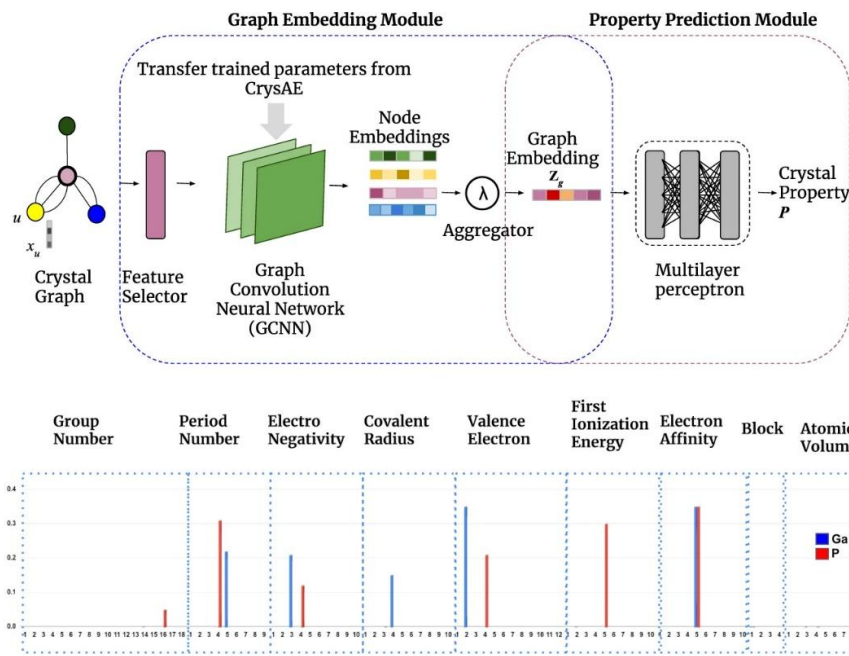

Figure 20. a) The architecture of CrysXPP, which is capable of producing explainable results, as seen in (b) the bar chart of features affecting the band gap of GaP crystal. Reproduced with permission [306]. Copyright 2022, Springer Nature.

### 7.2.2 Few-Shot Learning (FSL)

As mentioned, high-quality data in materials science which are complete with proper labels are scarce. This issue is exarcebated when we look at experimental data, which unlike computationally-produced data, is plagued with issues stemming from the different experimental equipments and variable environments. Therefore, the few-shot learning (FSL), which specifically targets situation where data is limited, has enticed materials scientists, especially the experimentalists. There are several approaches to FSL, as discussed in [309, 310], such as metric-based, optimization-based, and model-based approach. FSL is still a relatively young and unrefined method, but it has already attracted a lot of attentions. FSL has been implemented in the prediction of molecular properties [311, 312], classification of space group from electron backscatter diffraction (EBSD) data [313], and segmenting electron microscopy data [314].

## 8 Machine Learning Tasks for Material Science

This section will discuss the coverage of materials science tasks that ML tools have been utilized to assists in tackling. The common ML tasks in material science often coincide with the traditional ML tasks, which have been extensively studied and optimized. The tasks of inference of material property given structural and compositional data, generative modelling from a latent representation of desired properties, and the generation of DFT functionals, are analogous to the tasks that ML has traditionally performed well, including object classification, image and text generation using text cues, and natural language processing (NLP).

### 8.1 Potentials, Functionals, and Parameters Generation

Traditionally, the XC functionals used in DFT are generated through mathematical approaches, guided by empirical data, such as the Perdew-Zunger exchange, with the exact XC functional remains elusive. The search for an improved XC functional above the currently popular GGA on the Jacob's ladder of Perdew [315] is desirable. The techniques of ML have been started to be utilized in the generation of new XC functionals [316-320], with the aim of improving the calculated accuracy while maintaining the efficiency. Transferability remains a huge challenge, which will need a huge and diverse dataset to achieve.

The potentials and force fields used in molecular dynamics (MD) are critical in determining the reliability and accuracy of the output [321]. MD that does not involve first-principles approach but rather fixed potentials are in general less accurate the Ab Initio MD (AIMD), but they can be applied on a large system and long time scale, where AIMD is too costly. As such, one would hope that the standard MD can bring about results similar to AIMD. Developed in 2017, DeePMD [322] accurately reproduced the water model obtained from DFT. The same team developed an open-source tool for the on-the-fly generation of MD potentials, known as DP-GEN [323], available on available on https://github.com/deepmodeling/dpgen. In 2020, the team won the ACM Gordon Bell Prize for the DeePMD work, as it can be scaled efficiently on the best HPC. Similar works have also been carried out by other teams [324, 325].

Another material modeling technique is the Density Functional Tight Binding (DFTB) method, which is less computational expensive than DFT-based first principles calculations. Efforts have been carried out on applying ML to obtain the TB parameters [252, 326].

### 8.2 Screening of Materials

There are many m-ethods to compress design space. To name few, one could train a model that predicts material property given material information or performs ML guided simulation of new materials to predict material behaviour under certain circumstances. High-throughput screening eliminates most potential materials without actually performing actual experiments to verify their property and provide experimentalists with a minimal set of candidate materials to try out. Pivoting to a generative model perspective, one could also specify material properties and generate stable materials that are likely to fit the specified property [327-329].

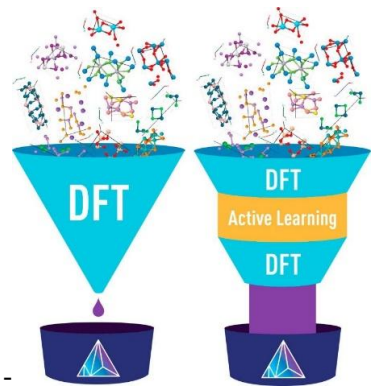

Figure 21: High-Throughput Screening with Learnt Interatomic Potential Embedding from [330]. With the integration of active learning and DFT in the screening pipeline, the throughput efficiency or the quality of the output obtained from calculation can be improved. Reproduced with permission [330]. Copyright 2019, Elsevier.

There are many properties that are of interests, including band gaps, bulk and shear moduli, crystal structures, conductivity, and topological states, as discussed in details in [216]. These properties are usually computed via DFT, which could be computationally expensive depending on the system setup. Properly setup and trained ML models can produce DFT-level accuracy properties predictions, while at far lower computational time. Isayev et al. [232] managed to obtain prediction at 0.1s for each structure, which amounts to 28 million structure in a day, as pictured in Figure 21. However, a well-trained and fully-transferable ML model requires the existence of high-quality large database and heavy computational power to optimize the model.

## 8.3 Novel Material Generation

The latent representations of common desired properties are of high interests among the community [331]. Based on the learnt latent representation, we can generate structure with similar desired properties at will. This is often carried out using the generative models, such as the GAN and VAE, as demonstrated in the work of Dong et al. [332] (see Fig. 22) and Pathak et al. [333].

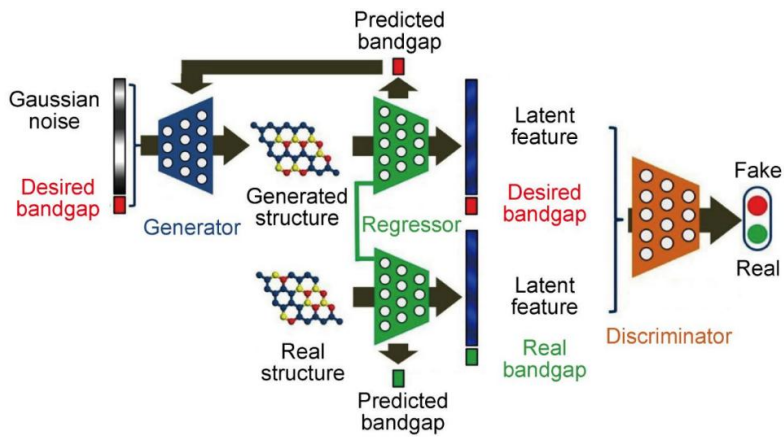

Figure 22: Schematics of generative adversarial network used in Dong et al. [332]. Reproduced with permission [332]. Copyright 2022, Elsevier.

## 8.4 Imaging Data Analysis

There are many imaging methods to capture the structure and fingerprints of a material experimentally at the atomic level, such as X-ray Diffraction (XRD), Fourier Transform Infrared Spectroscopy (FTIR), Atomic Force Microscopy (AFM), Transmission Electron Microscopy (TEM), and many others. Typically, they require laborious human interpretation to understand the meaning of the signals and whether they are due to noises and errors. This can be remedied with the help of ML, and thus far ML has been applied to:

- Identify symmetry and space group from XRD [334].

- Discover hidden atomic scale knowledge from XRD [297]

- Identify functional group in gaseous organic molecule [335]

- Analyze patterns and feature in AFM images, including domain wall and grain boundaries. [336]

- Quantify nanoscale forces in dynamic AFM [337]

- Perform structural analysis and reconstruction in TEM [338]

- Identify chemistry and processing history from microstructure image [298]
- Characterize and analyze mechanical behaviour in microstructure image [339]

## 8.5   Natural Language Processing of Material Science Literature

Nature Nature language processing (NLP) refers to the ability of computer algorithm to understand spoken and written langua ge. This technology has seen explosive development, with the recent GPT-3 [340] and now GPT-4 [341] models making strides not just academically, but used in just about everywhere. The trained AI models are able to hold realistic conversation with humans, take standardized exams [342], write codes in various programming language, and so on.

Most of the published results on materials science are not stored in a centralized database, which hinders the overall effort of ML applications. The NLP techniques can help this by scraping information from published literatures, such as the materials structure and properties. Tshitoyan et al. [343] demonstrated that an unsupervised learning modethat can capture complicated underlying knowledge from the literature and recommend materials for functional applications before their actual discovery. NLP also can help hypothesis formation and provide knowledge on the current trends in the field [344]. NLP methods can also serve as an efficient knowledge extractor from vast amount of material science literature, making the literature review process more efficient and thorough for researchers [345].

# 9 Perspectives on the Integration of Machine Learning in Materials Science

In the following we will list perspectives on the integration of machine learning in materials science with materal science point of view and with machine learning point of view, respectively.

## 9.1 Perspectives from Machine Learning Viewpoint

As ML techniques and ideals become ever more prevalent, we believe algorithmic templates and ML ideas will eventually become either the target modes of computation or the mode of guidelines which decides the permutation to which areas of material science garner attention and gain resources. Machine translation has evolved from a rule-based coupled with statistical model to a very data-driven approach, and researchers are discussing the translation task with less and less reference to a specific source and target languages, pivoting towards advancing mode of computation for the task as a whole.

### 9.1.1 More Deep Integrations

We might also observe the trend of attempting to learn descriptors for parts of complex systems with ML models to be either more computationally efficient or more human interpretable or editable. Instead of scientists attempting to describe a system with equations from first principles, ML models can help scientists discover a better set of descriptors for systems across all datasets. For example, descriptors could be descriptors for input data (atomistic information/reaction space) or labels (crystal structures). These discoveries can hugely impact physics and chemistry theory [346], experiments and research methods. Physics-informed neural networks [347] can both improve neural network performance and physics research efficiency.

Increasingly, it could be more and more about the mapping of descriptors. We can imagine that with the emergence of more sophisticated models, it is possible to advance a particular segment of study in material science, such as polymer design, by completing a well-defined sophisticated task with a model/model of computation, where the lack of relevant databases will limit its advancement. Tracing the development of computer science, sophisticated models which perform generic tasks in material science well will again be integrated into a giant multi-purpose model much like a generic processing chip, to which we can prompt for insights which was previously only gained by human experimentation at a much slower rate. ML models will bring material scientists closer to the many possibilities already inherent in big data itself, allowing us to explore and exploit the possibilities with greater efficiency. The task material scientist will be able to complete with the help of machine learning will become more integrated and sophisticated, from the screening of material to the design of material as a complete task. Then with the design of material as an atomic/primitivetransaction, we will be able to come out with new science on top of the material design as a whole.

### 9.1.2 Systematic Generalization

In our stride towards autonomous general intelligence (AGI), researchers have drawn many parallels and inspirations from neuro-sciences [348] and how humans learn and teach each other to develop models which better generalizes to novel situations and objects well. We expect a body of material science knowledge and ideas to become generalized and accessible to other fields, conveyed by advanced models in the future, where we can generalize or verbalize properties of imagined materials or predict performance of material in novel situations with high accuracy with its formal deduction process generated by models. We can also observe the interaction, cooperation or contradiction between bodies of materials science knowledge for novel materials and circumstances, and perform research on the intersection of bodies of knowledge with more depth and rigor. ML models can also learn to identify potential directions for exploration, come up with a comprehensive experimentation plan and collaborate with human researchers as a navigation co-pilot. The novel direction identified will be novel and comprehensive because models can learn from passive observations of a large material science literature, its publication trend [349] and insight analysis of researchers.

### 9.1.3 Huge Computational Models

With the development of reaction environment models, one might also reasonably expect reinforcement learning game-play learning to learn an agent policy for a material, i.e. to first learn a material behaviour that is desirable for a particular purpose, followed by an automatic search/generation of material which suits the specification. In general, the compounding of learning methods to get a solution for an even more vaguely defined objective but more analyzable process for that solution results in a human-verifiable solution for large or vague problems. Moreover, the increasing synthesizability or explainability of the solution to vague problems will help material scientists navigate methods for solving huge overarching/generic problems with more finesse, evolution of subject through large models [350].

Model of computation might become the common language of material scientists and researchers from other fields. Task definition might become the lingua franca or the leading cause of concern for ML practitioners in material science. This broader definition of material science might then, in turn, propel the advancement of machine learning. In general, the barrier to entry to both advanced material science and advanced ML will be lowered, allowing more experts from other fields or individuals to contribute their efforts and ideas to the development of both fields.

Mechanisms in quantum ML will become readily-available to be integrated with quantum physics, chemistry and subsequently material science. As classical-quantum hybrid infrastructure and architectures [351] become more available, quantum learning for material science might incorporate mechanisms of both quantum computing and quantum analysis of materials as primitives. This trend is expected to speed up the inter-disciplinary mixing of these fields from both engineering and theoretical grounds.

The resulting phenomenon is the emergence of an ever more integrated huge ML model, a Super Deep-Learning Model, which will tackle most if not all of the most fundamental underlying problems in material science; it will integrate fundamental engineering ideas from computer science with domain-invariants of material science, which is designed to perform well for various tasks on both super-computing facilities, quantum or otherwise, and on limited resources devices, [352] scalable yet robust. Moreover, by integrating the best training and privacy practices from ML software and hardware development experience, future material scientists can quickly expect robust material science downstream models running smoothly and reliably as an application on widely available and portable devices like a cellphone.

## 9.2 Perspectives from Material Science Viewpoint

Currently, one of the biggest challenges is the availability of high-quality data. The increasing number of research groups adapting the open data approach and the growing availability of internet of things (IoT) devices will solve this problem, albeit gradually. We have also discussed several possible methods to overcome the issues, which is the advancement of small training sample ML algorithms, such as transfer learning and few-shot learning algorithms will also be one of the possible solutions to this issue.

### 9.2.1 Theoretical and Computational Materials Science

The various computational techniques in materials science, such as DFT, molecular dynamics in its various forms of molecular dynamics (MD), monte-carlo methods, and density functional tight binding method, has started to benefit from the application of ML and will continue to do so in a dramatic manner.

As of now, the Kohn-Sham DFT remains a reliable and popular method for determining various material properties. However, the accuracy of DFT calculations heavily relies on the quality of the approximations employed, such as the exchange-correlation (XC) functional. The search for improved approximations, including exact functionals, using ML has only recently commenced. Another area for improvement in DFT is reducing computational costs. Recently, ML-refined numerical techniques have emerged that offer faster speeds compared to their traditional counterparts [353-355]. It is hoped that these advancements can eventually be applied to accelerate DFT computations.

The integration of ML into MD, exemplified by methods like DeePMD, has demonstrated the potential to achieve DFT-level accuracy while maintaining the computational efficiency of classical MD. This breakthrough opens up new possibilities for conducting accurate calculations in ab initio molecular dynamics (AIMD) on extremely large systems (with over 100 million atoms) or over very long timescales (beyond 1 nanosecond) [333, 355]. By enabling adequate sampling of phase space, these advancements enable more comprehensive investigations across various applications, including (electro-)catalysis, sensors, fabrication, drug interactions, and more.

### 9.2.2 Experimental Materials Science

The availability of a vast number of predicted materials with desired properties is highly advantageous for experimentalists. With a large number of possible candidates, the experimentalist can focus on the materials that can be synthesized and tested on available facilities and equipments. Additionally, the automated learning of the fabrication parameters and conditions are on the rise recently [356-359]. The advancement of MD will also enable comprehensive simulations of fabrication process and finds out the best experimental conditions for successful synthesis of new materials [360-361]. Furthermore, the analysis of the data, a highly time-consuming and laborious task is being increasingly supported by ML algorithms. The implementation of on-the-fly accurate inference mechanism of experimental data will increase the producitivity and efficiency of fabrication process, enabling experimentalists to determine if the samples have been fabricated successfully and move on to the next attempt quickly.

### 9.2.3 Coupling of Data-driven Discovery With Traditional Techniques

The art of tailoring and creating materials with desired properties -- materials engineering, includes techniques such as defect engineering [361-363], doping [44, 362, 364-369], fluorinating [370] or alloy engineering [30-34, 371] or salt engineering [35] by varying composition, strain engineering [11, 16, 26, 36, 170, 372-374] by applying mechanical load (such as hydro-static pressure [36, 375-379] or directional stress), and interfacial engineering [40, 42-46] by choosing different materials for forming interface with novel or exotic properties. These methods have been demonstrated to be very useful for tuning materials properties or creating new materials with advantageous properties. In quantum materials [14, 28, 37, 380], including strongly correlated functional materials and superconducting materials, the charge-spin-orbital engineering plays a crucial role in controlling the quantum behaviour.

The availability of advance X-ray scattering and electron scattering techniques [14, 28, 191, 381, 382] such as synchrotron radiation and electron microscropy, and advancing nanotehnologies and simulation methods, have led to an increasingly growing amount of high quality experimental or simulated data, available for data-driven discovery and data mining. The integration of data-driven discovery with traditional techniques are expected to play an increasingly important role in materials science research at various length and time scales, ranging from microscropic scale to macroscropic scale.

## 10 Conclusion

In conclusion, this review briefly introduced basic concepts and history of machine learning, and provided detailed information of coupling between machine learning and materials science in fundamental and technical perspectives. The nuances of machine learning, from the descriptors to the various algorithms, have been discussed in the context of materials science. Besides, this review also covered the tasks or issues in materials scicence that has been tackled with the use of machine learning. We also discussed our vision for the future of materials science as the field matures with the integration of machine learning, which will be drastically different from what we know today.

### Acknowledgements

This research was supported by the Ministry of Higher Education Malaysia through the Fundamental Research Grant Scheme (FRGS/1/2021/STG05/XMU/01/1).